\documentclass[10pt,twocolumn,letterpaper]{article}

\usepackage{iccv}
\usepackage{times}
\usepackage{epsfig}
\usepackage{graphicx}
\usepackage{amsmath}
\usepackage{amssymb}
\usepackage{booktabs}
\usepackage{multirow}
\usepackage{color}
\usepackage{colortbl}
\usepackage{verbatim}
\newtheorem{Thm}{Theorem}

\usepackage{mathrsfs}
\usepackage{MnSymbol}
\usepackage{utfsym}
\usepackage[title]{appendix}

\usepackage{algorithm}
\usepackage[noend]{algpseudocode}
\usepackage{algorithmicx}
\usepackage{listings}

\usepackage[pagebackref=true,breaklinks=true,letterpaper=true,colorlinks,bookmarks=false]{hyperref}
\usepackage[capitalize]{cleveref}
\usepackage[accsupp]{axessibility} 
\crefname{section}{Sec.}{Secs.}
\Crefname{section}{Section}{Sections}
\Crefname{table}{Table}{Tables}
\crefname{table}{Tab.}{Tabs.}

\iccvfinalcopy 

\def\iccvPaperID{1990} 

\ificcvfinal\fi

\begin{document}

\title{Progressive Content-aware Coded Hyperspectral Compressive Imaging}

\author{
Xuanyu Zhang$^{1}$, Bin Chen$^{1}$, Wenzhen Zou$^{2}$, Shuai Liu$^{3}$, Yongbing Zhang$^{2}$, Ruiqin Xiong$^{4}$, Jian Zhang$^{1}$\\
$^{1}$Peking University Shenzhen Graduate School $\quad^{2}$Harbin Institute of Technology (Shenzhen) \\
$^{3}$Tsinghua University $\quad^{4}$Peking University \\
}

\maketitle
\ificcvfinal\thispagestyle{empty}\fi

\let\thefootnote\relax\footnotetext{Corresponding author: Jian Zhang (zhangjian.sz@pku.edu.cn).}
\begin{abstract}
Hyperspectral imaging plays a pivotal role in a wide range of applications, like remote sensing, medicine, and cytology. By acquiring 3D hyperspectral images (HSIs) via 2D sensors, the coded aperture snapshot spectral imaging (CASSI) has achieved great success due to its hardware-friendly implementation and fast imaging speed. However, for some less spectrally sparse scenes, single snapshot and unreasonable coded aperture design tend to make HSI recovery more ill-posed and yield poor spatial and spectral fidelity. In this paper, we propose a novel \textbf{P}rogressive \textbf{C}ontent-\textbf{A}ware \textbf{CASSI} framework, dubbed \textbf{PCA-CASSI}, which captures HSIs with multiple optimized content-aware coded apertures and fuses all the snapshots for reconstruction progressively. Simultaneously, by mapping the Range-Null space Decomposition (RND) into a deep network with several phases, an RND-HRNet is proposed for HSI recovery. Each recovery phase can fully exploit the hidden physical information in the coded apertures via explicit $\mathcal{R}$$-$$\mathcal{N}$ decomposition and explore the spatial-spectral correlation by dual transformer blocks. Our method is validated to surpass other state-of-the-art methods on both multiple- and single-shot HSI imaging tasks by large margins. 
\end{abstract}

\section{Introduction}
\label{sec:intro}
Hyperspectral images (HSIs) embody rich spectral bands and detailed information than the normal RGB images, which have unprecedented demand in recent years \cite{yuan2021snapshot,arce2013compressive,cao2016computational,he2021fast}. Inspired by the compressive sensing (CS) \cite{zhao2016video, zhang2014image, li2022d3c2, chen2022content, zhang2023physics, wang2022zero, wang2022gan, hu2023dear}, CASSI system aims to utilize 2D detector to capture 3D hyperspectral scenes. Due to the merits of fast acquisition speed, low cost, and high data throughput, it has played an indispensable role in a wealth of applications, such as remote sensing, object detection, super-resolution, and medical diagnosis \cite{zhao2015sub,ojha2015spectral,melgani2004classification,lei2019spectral,xie2019structure,ding2020hyperspectral,wei2019medical,akbari2010detection}. 
\begin{figure}[t]
	\centering
	\includegraphics[width=1\linewidth]{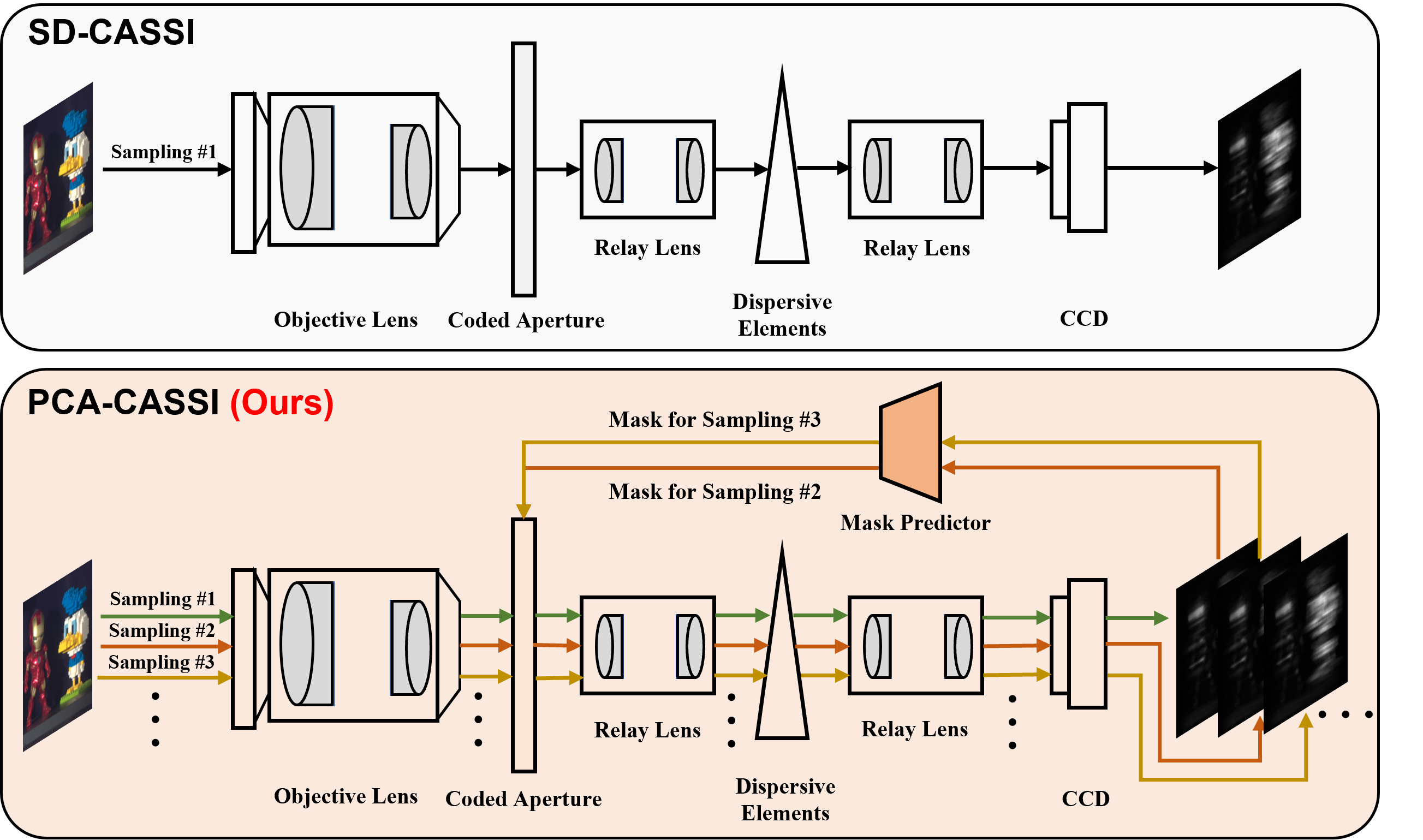}
	\vspace{-10pt}
	\caption{Illustration of SD-CASSI and the proposed PCA-CASSI system. SD-CASSI compresses the HSI via a single coded aperture and tends to cause excessive information loss. However, our PCA-CASSI captures the same HSI with multiple content-aware coded apertures (produced by the mask predictor) progressively, which significantly increases the information throughput.}
	\label{intro}
	\vspace{-15pt}
\end{figure}
However, for some specific applications, the information retained in a single snapshot is inadequate. For instance, microscopic imaging \cite{bullen2008microscopic} has extremely high demands on the textures and details of the HSIs. To guarantee the accuracy of recovered images, capturing the same scene with multiple shots is necessary and imperative. Simultaneously, with the development of sampling devices, the improved digital micromirror device (DMD) \cite{wu2011development} and CCD sensors allow the imaging system to record multiple snapshots and alter the coded patterns with a limited increase in acquisition time. 

Although previous works such as MS-CASSI \cite{kittle2010multiframe} have explored the potential of multiple shots, there are still several bottlenecks to be solved. \textbf{First}, how to design optimal multiple-coded apertures is the key to the enhancement of multiple snapshot imaging systems. We focus on the following two aspects: \textbf{1)} The coded apertures are expected to exhibit excellent anisotropy and complementarity. The complementary coded apertures promote the fusion and interaction of the multiple snapshots, otherwise they may tend to interfere with each other and undermine the beneficial information; \textbf{2)} Furthermore, the coded apertures are supposed to be adjusted contextually. Prior information from the previous shots enables coded apertures to perceive the content of HSIs, thus increasing the flexibility and performance of imaging systems. To the best of our knowledge, these two aspects remain under-investigated.
\textbf{Second}, existing reconstruction networks generally focus on the single disperser CASSI system (SD-CASSI). The practical solution for multiple-shot reconstruction is deficient. Meanwhile, how to utilize the coded aperture to retain the range space and recover the null space of HSIs has been ignored.

To solve the above-mentioned issues, a novel \textbf{P}rogressive \textbf{C}ontent-\textbf{A}ware \textbf{C}oded \textbf{A}perture \textbf{S}napshot \textbf{S}pectral \textbf{I}maging framework, dubbed \textbf{PCA-CASSI}, is proposed for progressive content-aware sampling and accurate multiple-shot reconstruction (Fig.~\ref{intro}). From hardware perspective, PCA-CASSI can be easily implemented from the original SD-CASSI \cite{wagadarikar2008single} hardware with little modification. From algorithm perspective, PCA-CASSI is composed of the light-weight recovery network and mask predictor. Furthermore, an \textbf{R}ange-\textbf{N}ull space \textbf{D}ecomposition unfolding \textbf{H}yperspectral \textbf{R}econstruction \textbf{Net}work (RND-HRNet) is proposed to fuse all the coded measurements adaptively. Overall, our contributions are summarized as:
\begin{itemize}
	\item A novel ``\textcolor{red}{Encoder} + \textcolor{blue}{Decoder}'' framework, dubbed PCA-CASSI, is proposed to capture HSIs contextually and reconstruct them progressively. Noted that the ``\textcolor{red}{Encoder}'' corresponds to the HSI sampling while the ``\textcolor{blue}{Decoder}'' actually indicates the HSI reconstruction.
	
	\item A progressive content-aware sampling strategy is proposed to perceive HSI contents and optimize the coded aperture in the current shot from the measurement in the previous shot.
	
	\item An $\mathcal{R}$$-$$\mathcal{N}$ decomposition-inspired network, dubbed RND-HRNet, is presented to reconstruct HSIs, which utilizes the range-null space decomposition module (RNDM) to refine the null space of the HSIs iteratively and adopts the spectral spatial fusion module (SSFM) to exploit the non-local spatial-spectral information.
	
	\item Experiments exhibit that our method outperforms other state-of-the-art methods on both multiple- and single-shot HSI reconstruction tasks by large margins.
\end{itemize}

\section{Related Works}
\label{sec:related_works}
\subsection{Hyperspectral Imaging Systems}
Conventional hyperspectral cameras usually make a trade-off between temporal and spectral resolution. Inspired by the theory of compressive sensing (CS), CASSI aims to capture hyperspectral images (HSIs) within a snapshot time. In what follows, we will review some representative works on imaging optical path and coded aperture design.
\begin{figure*}[t]
	\centering
	\includegraphics[width=1\linewidth]{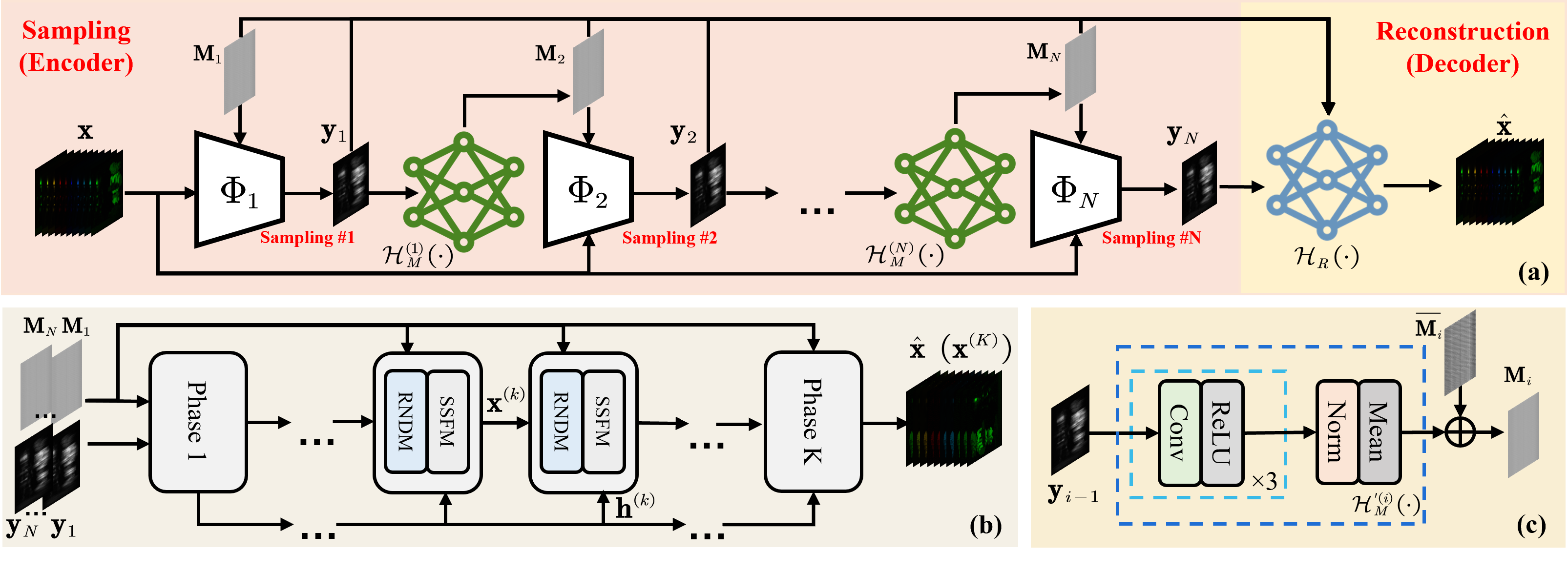}
	\caption{Illustration of the proposed PCA-CASSI. (a) overall architecture, (b) $\mathcal{H}_R(\cdot)$, (c) $\mathcal{H}_M^{(i)}(\cdot)$. $\mathcal{H}_R(\cdot)$ takes multiple snapshots $\{\mathbf{y}_i\}_{i=1}^{N}$ and coded apertures $\{\mathbf{M}_i\}_{i=1}^{N}$ to generate the reconstructed HSI $\hat{\mathbf{x}}$, which includes $K$ RND-inspired recovery phases. It is also adapted to single-shot reconstruction. $\mathcal{H}_M^{(i)}(\cdot)$ takes the snapshot $\mathbf{y}_{i-1}$ from previous shot to predict the coded aperture $\mathbf{M}_i$ in the current shot.}
	\label{pca-cassi}
	\vspace{-10pt}
\end{figure*}

{\noindent\textbf{Imaging Optical Path:}} The most representative CASSI system \cite{wagadarikar2008single} utilized a single disperser to encode spatial and spectral information. To improve the imaging performance and information throughput, Kittle \emph{et al.} \cite{kittle2010multiframe} captured the same hyperspectral scenes via varied coded apertures. In addition, by temporally aligning the single-shot CASSI system with an RGB camera, Wang \emph{et al.} \cite{wang2015dual} provided multi-modal supervision and complementary information for HSI reconstruction. Recently, Lin \emph{et al.} \cite{lin2014dual, lin2014spatial} utilized two high speed spatial light modulators to realize dual-optical coding. Although the optical path design of CASSI has achieved remarkable results, ensuring high compression ratio and excellent image quality is still a bottleneck. 

{\noindent\textbf{Coded Aperture Design:}} Wu \emph{et al.} \cite{wu2011development} first introduced the digital micromirror device (DMD) into the imaging systems to realize flexible and fast dynamic coding. Benefiting from the rapid response time of DMD, the theory of Restricted Isometry Property (RIP) \cite{eldar2012compressed} was introduced to guide the coded aperture optimization. To make the coded aperture retain enough useful information and remove redundancy, Zhang \emph{et al.} \cite{zhang2022herosnet} combined the coded aperture optimization and HSI reconstruction with an united network to achieve efficient mask optimization. However, as the complex scene varies, designing adaptive, content-aware and task-driven coded apertures is worthy of exploration.

\subsection{HSI Reconstruction Methods}
\noindent\textbf{Model-based Methods:} The traditional model-based methods employ the regularization term inspired by the image prior to solve the ill-posed inverse problem iteratively with widely-used optimization frameworks, such as ISTA \cite{bioucas2007new}, GAP \cite{yuan2016generalized} and so on \cite{ma2019deep, tan2015compressive}. Simultaneously, to improve the representation ability of the model, TV \cite{yuan2016generalized}, GMM \cite{yang2014video, yang2014compressive}, image sparsity representation \cite{wang2016adaptive} and rank minimization \cite{liu2018rank} are embedded into the above optimization frameworks to provide prior information. Although these methods produce decent results in specific scenarios, it is difficult to design suitable hand-crafted priors for all scenes.

\noindent\textbf{Deep Learning-based Methods:} Owing to the powerful representation capabilities of deep networks, learning-based HSI reconstruction methods \cite{miao2019net} have received increasing attention. To improve the accuracy and perceptual quality, residual blocks \cite{wang2021new}, spatial-spectral attention modules \cite{meng2020end}, long-short-term memory units \cite{cheng2022recurrent} and Fourier domain constraint \cite{hu2022hdnet} are embedded into the structure of convolution neural networks (CNNs). Although they can capture the local image cues, CNNs fail to exploit the global correlation and long-range dependencies. Recently, thanks to the wide application of vision transformers, spatial-wise and channel-wise transformers have been incorporated into the multi-scale encoder-decoder architectures \cite{cai2022mask, cai2022coarse}. To alleviate the poor interpretability of these end-to-end networks, some researchers have attempted to combine optimization algorithms and deep network priors, such as plug-and-play frameworks \cite{yuan2020plug, zheng2021deep, meng2021self}. Meanwhile, deep unfolding methods \cite{wang2019hyperspectral,wang2020dnu,zhang2021learning, caidegradation, huang2021deep, zhang2022herosnet} have been prevalent for its exquisite design and powerful performance. Although these methods have achieved great success, few attention has been paid to the multiple-shot HSI reconstruction.

\section{Proposed Method}
\label{sec:method}
\subsection{Review of the CASSI system}
\label{sec:cassi}
In CASSI, the 3D hyperspectral cube is first modulated via a coded aperture and then dispersed via a dispersive element (Fig.~\ref{intro}). Given a HSI sequence $\{\mathbf{X}_{i}\}_{i=1}^{C}$$\in $$\mathbb{R}^{H \times W}$, each frame is modulated via a coded aperture $\mathbf{M}$$ \in $$\mathbb{R}^{H \times W}$:
{\setlength\abovedisplayskip{0.1cm}
\setlength\belowdisplayskip{0.1cm}
\begin{equation}
	\mathbf{X}_{i}^{\prime}=\mathbf{M} \odot \mathbf{X}_{i},
	\label{eq1}
\end{equation}}where $\mathbf{X}_{i}^{\prime}$ is the $i^{th}$ modulated HSI frame and $\odot$ denotes the Hadamard product. The modulated HSI frames $\{\mathbf{X}_{i}^{\prime}\}_{i=1}^{C}$$ \in $$\mathbb{R}^{H \times W}$ with different wavelengths are then shifted spatially and summed in an element-wise manner, leading to a coded measurement:
{\setlength\abovedisplayskip{0.1cm}
\setlength\belowdisplayskip{0.1cm}
\begin{equation}
	\mathbf{Y}(m,n)=\sum_{i=1}^{C} \mathbf{X}_{i}^{\prime}(m, n+d_i)+\mathbf{N},
	\label{eq2}
\end{equation}}where $(m, n)$ is the spatial coordinates, and $d_i$ denotes the shifting distance of the $i^{th}$ channel. $\mathbf{N}$$\in $$\mathbb{R}^{H \times (W+C-1)}$ and $\mathbf{Y}$$\in$$\mathbb{R}^{H \times (W+C-1)}$ denote the noise and the coded measurement, respectively. The vectorized form of CASSI is:
{\setlength\abovedisplayskip{0.1cm}
\setlength\belowdisplayskip{0.1cm}
\begin{equation}
	\mathbf{y}=\mathbf{\Phi} \mathbf{x}+\mathbf{n},
	\label{eq3}
\end{equation}}where $\mathbf{x}$, $\mathbf{y}$ and $\mathbf{n}$ denote the vectorized form of $\mathbf{X}$, $\mathbf{Y}$ and $\mathbf{N}$, respectively. $\mathbf{\Phi}$ represents the sensing matrix.

The above imaging system can be extended to multiple shots. For instance, if we capture the same scene $\mathbf{X}$ via $N$ multiple different code apertures $\{\mathbf{M}_{i}\}_{i=1}^{N}$, every shot can be treated as an implementation of the CASSI system. Hereby, the imaging scheme can be formulated as follows:
\vspace{-10pt}
{\setlength\abovedisplayskip{0.1cm}
\setlength\belowdisplayskip{0.1cm}
\begin{equation}
\left[\mathbf{y}_1^{\top}, \mathbf{y}_2^{\top}, \ldots, \mathbf{y}_N^{\top}\right]^{\top} = \left[\mathbf{\Phi}_1^{\top}, \mathbf{\Phi}_2^{\top}, \ldots, \mathbf{\Phi}_N^{\top}\right]^{\top}\mathbf{x} + \mathbf{n},
\label{eq4}
\end{equation}
}where $\mathbf{y}_{i}$ and $\mathbf{\Phi}_{i} = \operatorname{Mask2Mat}(\mathbf{M}_i)$ denote the observed HSI measurement and sensing matrix of the $i^{th}$ shot, respectively. $\operatorname{Mask2Mat}(\cdot)$ transforms the physical mask $\mathbf{M}_i$ to its equivalent sensing matrix form $\mathbf{\Phi}_i$. The degradation process of multiple-snapshot compressive imaging remains a linear model and can still be expressed as $\mathbf{y} = \mathbf{\Phi}\mathbf{x}+\mathbf{n}$.

\subsection{Overview of the Proposed PCA-CASSI}
As illustrated in Fig.~\ref{pca-cassi}, PCA-CASSI is a novel ``Encoder + Decoder" framework, which aims to compress HSIs by various content-aware optimal masks to increase the information throughput and fuse all the captured snapshots for HSI restoration cooperatively. Note that following \cite{cai2022coarse, cai2022mask}, this work just focuses on hyperspectral image sampling and restoration, and our contributions may be extensible to other imaging systems like \cite{wagadarikar2008single,  kittle2010multiframe, Zhang_2019_ICCV}. For software algorithm, it is composed of two sub-modules, namely the mask predictors $\{\mathcal{H}_M^{(i)}(\cdot)\}_{i=1}^{N}$ and reconstruction network $\mathcal{H}_R(\cdot)$. The mask predictors $\{\mathcal{H}_M^{(i)}(\cdot)\}_{i=1}^{N}$ are capable of generating the optimal mask contextually in the current shot from the measurement in the previous shot, which is formulated as:
{\setlength\abovedisplayskip{0.1cm}
\setlength\belowdisplayskip{0.1cm}
\begin{equation}
    \mathbf{M}_{i}= \mathcal{H}_M^{(i)}\left(\mathbf{y}_{i-1}\right), \quad i\in\{2,3,\cdots\} ,
    \label{eq6}
\end{equation}}where $\mathbf{M}_{i}$ denotes the mask in the $i^{th}$ shot and $\mathbf{y}_{i-1}$ represents the measurement in the ${(i-1)}^{th}$ shot. $\mathbf{M}_1$ is a learnable parameter. We will elaborate this module in Sec.~\ref{sec:pca}. Simultaneously, $\mathcal{H}_R(\cdot)$ is designed to integrate all the coded snapshots based on the $\mathcal{R}$$-$$\mathcal{N}$ Decomposition (RND) theory, which is formulated as follows:
{\setlength\abovedisplayskip{0.1cm}
\setlength\belowdisplayskip{0.1cm}
\begin{equation}
    \hat{\mathbf{x}}=\mathcal{H}_R\left(\mathbf{y}_1, \mathbf{M}_1, \mathbf{y}_2, \mathbf{M}_2, \ldots, \mathbf{y}_{N}, \mathbf{M}_{N}\right),
    \label{eq7}
\end{equation}}where $\hat{\mathbf{x}}$ denotes the reconstruction result. We will elaborate it in Sec~\ref{sec:rnd}. For hardware realization, the proposed PCA-CASSI can be easily implemented via the acquisition devices of SD-CASSI \cite{wagadarikar2008single}. Owing to the rapid response time of 2D sensors and the fast prediction speed of $\mathcal{H}_{M}^{(i)}(\cdot)$, the increased time caused by multiple shots is indeed limited. Considering its superior performance, it is still worth sacrificing a little sampling speed for higher imaging performance in some applications. As illustrated in Tab.~\ref{syscomp}, our framework is distinguished from some previous works. Different from MS-CASSI \cite{kittle2010multiframe} which utilizes parallel multiple coded apertures in the optical path, the proposed PCA-CASSI can achieve progressive sampling sequentially and be easily extended to any number of shots. Different from DCD \cite{YingFu2021CodedHI, Zhang_2019_ICCV}, which uses dual cameras to capture snapshots and grayscale images simultaneously, our framework requires only one camera and is more flexible with respect to the coded apertures. Hence, the proposed PCA-CASSI is hardware-friendly in system implementation, content-aware in coded aperture design, and effective in HSI recovery.

\begin{table}[]
\caption{High-level functional comparison among three representative imaging systems and our proposed PCA-CASSI.}
\resizebox{0.48\textwidth}{!}{
\begin{tabular}{c|c|c|c|c}
	\toprule[1.5pt]
          \cellcolor[HTML]{F2F2F2} Method& \cellcolor[HTML]{F2F2F2}SD-CASSI~\cite{wagadarikar2008single}& \cellcolor[HTML]{F2F2F2}DCD~\cite{Zhang_2019_ICCV} & \cellcolor[HTML]{F2F2F2}MS-CASSI~\cite{kittle2010multiframe} & \cellcolor[HTML]{F2F2F2}\textbf{PCA-CASSI} \\ \midrule[0.75pt]
Single CCD sensor  &$\usym{2713}$ & $\usym{2717}$ &$\usym{2713}$ &$\usym{2713}$  \\
Multiple mask switch &$\usym{2717}$ &$\usym{2717}$ &$\usym{2713}$ &$\usym{2713}$ \\ 
Progressive sampling &$\usym{2717}$ &$\usym{2717}$ &$\usym{2717}$ &$\usym{2713}$  \\
Content-aware mask &$\usym{2717}$ &$\usym{2717}$ &$\usym{2717}$ &$\usym{2713}$  \\ \bottomrule[1.5pt]
\end{tabular}}
\label{syscomp}
\vspace{-15pt}
\end{table}

\subsection{Progressive Content-aware Sampling}
\label{sec:pca}
Apart from the fusion mechanism in the restoration process, coded aperture design plays a pivotal role in PCA-CASSI, which enables different measurements to contain complementary information and maintain excellent anisotropy. Considering that different HSIs have various spectral correlations and spatial sparsities, the coded measurement from the previous shot is utilized to optimize the coded aperture in the current shot. Thereby, the optimized coded apertures tend to perceive the photographed HSIs and make content-aware adjustments adaptively. As the shot epoch progresses, the optimized coded apertures tend to be more reasonable with more accurate reconstruction results.

Specifically, as shown in Fig.~\ref{pca-cassi} \textcolor{red}{(a)}, the hyperspectral scene $\mathbf{x}$ is firstly captured by the mask $\mathbf{M}_1$ to obtain the measurement $\mathbf{y}_1$. Then, the coarse compressed result $\mathbf{y}_1$ is adopted to furnish prior information and predict subsequent mask $\mathbf{M}_2$. Due to the guidance of $\mathbf{y}_1$, $\mathbf{M}_2$ is able to perceive the content of the HSIs and refine the coded pattern. Generally, the coded aperture $\mathbf{M}_i$ in the $i^{th}$ shot is generated from the measurement $\mathbf{y}_{i-1}$ via the mask predictor $\mathcal{H}_{M}^{(i)}(\cdot)$. To be noted, the optimized mask in each shot is required to reflect both the shared properties of the imaging system and the independent characteristics of each HSI. Hence, the optimized masks $\{\mathbf{M}_i\}_{i=1}^{N}$ are decoupled into two components, namely shared component and content-aware component. The shared components $\{\overline{\mathbf{M}_i}\}_{i=1}^{N}$ are learnable parameters and jointly optimized in the network. The content-aware component is derived from the previous measurement $\mathbf{y}_{i-1}$ via a lightweight deep module $\mathcal{H}_M^{\prime(i)}(\cdot)$, composed of three Conv-ReLU layers, a normalization layer and a mean layer (Fig.~\ref{pca-cassi} \textcolor{red}{(c)}). The normalization layer transforms all pixels to the interval $[$0, 1$]$. The mean layer calculates the channel-wise means of features to produce the masks. The above pipeline, corresponding to Eq.~\ref{eq6}, is implemented as:
{\setlength\abovedisplayskip{0.1cm}
\setlength\belowdisplayskip{0.1cm}
\begin{equation}
    \mathbf{M}_{i}= \mathcal{H}_M^{(i)}\left(\mathbf{y}_{i-1}\right)=\overline{\mathbf{M}_{i}}+\eta^{(i)} \cdot \mathcal{H}_M^{\prime(i)}(\mathbf{y}_{i-1}),
    \label{eq8}
\end{equation}}where $\eta^{(i)}$ denotes a learnable parameter to stabilize the network training. By means of our progressive content-aware sampling, each snapshot tends to capture complementary physical information and spectral features.

\subsection{Architecture of the Proposed RND-HRNet}
\label{sec:rnd}
The proposed RND-HRNet aims to recover the hyperspectral scene $\mathbf{x}$ from the snapshots $\{\mathbf{y}_i\}_{i=1}^{N}$ and use coded apertures $\{\mathbf{M}_i\}_{i=1}^{N}$ to guide the transmission of spectral-spatial features. To start with the noise-free degradation model $\mathbf{y}$$=$$\mathbf{\Phi}\mathbf{x}$, the HSI reconstruction is formulated as:
{\setlength\abovedisplayskip{0.1cm}
\setlength\belowdisplayskip{0.1cm}
\begin{equation}
	\hat{\mathbf{x}}=\arg \min _{\mathbf{x}} \frac{1}{2}\|\mathbf{y}-\mathbf{\Phi} \mathbf{x}\|_{2}^{2}+\lambda \psi(\mathbf{x}),
	\label{eq9}
\end{equation}}where $\mathbf{y}=\left[\mathbf{y}_1^{\top}, \mathbf{y}_2^{\top}, \ldots, \mathbf{y}_N^{\top}\right]^{\top}$ denotes the concatenation of all vectors of measurements. The first part is the data fidelity term, and the second part $\lambda \psi(\mathbf{x})$ denotes the regularization term. Following the traditional ISTA framework \cite{zhang2018ista, zhang2020optimization}, Eq.~\ref{eq9} can be solved by iterating between the following gradient descent and proximal mapping steps:
{\setlength\abovedisplayskip{0.1cm}
\setlength\belowdisplayskip{0.1cm}
\begin{equation}
	\mathbf{r}^{(k)}=\mathbf{x}^{(k-1)}-\rho \mathbf{\Phi}^{\top}(\mathbf{\Phi} \mathbf{x}^{(k-1)}-\mathbf{y}),
	\label{eq10}
\end{equation}}
\vspace{-2.5mm}
{\setlength\abovedisplayskip{0.1cm}
\setlength\belowdisplayskip{0.1cm}
\begin{equation}
	\mathbf{x}^{(k)}=\operatorname{prox}_{\lambda \psi}(\mathbf{r}^{(k)}),
	\label{eq11}
\end{equation}}where $\mathbf{r}^{(k)}$ and $\mathbf{x}^{(k)}$ denote the intermediate result and reconstruction image in the $k^{th}$ phase, respectively. $k$ and $\rho$ denote the number of ISTA iterations and the step size, respectively. The limitation of the above Eq.~\ref{eq10} is that gradient descent can only find an approximate sub-optimal solution to the data fidelity term in each iteration, which cannot strictly satisfy $\mathbf{y} \equiv \mathbf{\Phi}\mathbf{r}^{(k)}$. To tackle with the bottleneck of gradient descent, inspired by \cite{chen2020deep, DongdongChen2021EquivariantIL}, the $\mathcal{R}$$-$$\mathcal{N}$ Decomposition (RND) is introduced to explicitly maintain the consistency constraint of the data fidelity term. Given a sensing matrix $\mathbf{\Phi}$ and its pseudo inverse matrix $\mathbf{\Phi}^{\dagger}$, which satisfies $\mathbf{\Phi} \mathbf{\Phi}^{\dagger} = \mathbf{I}$, we have the following theorem for arbitrary HSIs:
\begin{Thm}[\cite{chen2020deep}]
$\mathcal{R}-\mathcal{N}$ Decomposition: Let $\mathcal{P}_r \triangleq \boldsymbol{\Phi}^{\dagger} \boldsymbol{\Phi}$ be the operator that projects the sample $\mathbf{x}$ from sample domain to the range of $\boldsymbol{\Phi}^{\dagger}$, and denote by $\mathcal{P}_n \triangleq\left(\boldsymbol{I}-\boldsymbol{\Phi}^{\dagger} \boldsymbol{\Phi}\right)$ the operator that projects $\mathbf{x}$ to the null space of $\mathbf{\Phi}$. Then $\forall \mathbf{x} \in \mathbb{R}^{H \times  W \times C}$, there exists the unique decomposition:
{\setlength\abovedisplayskip{0.1cm}
\setlength\belowdisplayskip{0.1cm}
\begin{equation}
    \mathbf{x} \equiv \mathcal{P}_r(\mathbf{x}) + \mathcal{P}_n (\mathbf{x}),
    \label{eq12}
\end{equation}}
\end{Thm}
where $\mathcal{P}_r(\mathbf{x})$ and $\mathcal{P}_n(\mathbf{x})$ respectively denote the range and null space of $\mathbf{x}$. The HSI reconstruction can be treated as solving these two components $\mathcal{P}_r(\mathbf{x})$ and $\mathcal{P}_n(\mathbf{x})$. Reconstructing the range space of the HSI signal is to ensure the data consistency with respect to measurement $\mathbf{y}$, while refining the null space of the HSI signal aims to remove artifacts and enrich details. Therefore, the core advantage of RND lies in its ability to improve perceptual quality while maintaining reconstruction fidelity. Furthermore, substituting $\mathbf{y}=\boldsymbol{\Phi}\mathbf{x}$ into Eq.~\ref{eq12}, we have the following RND for $\mathbf{x}$: 
{\setlength\abovedisplayskip{0.1cm}
\setlength\belowdisplayskip{0.1cm}
\begin{equation}
\mathbf{x} \equiv \mathbf{\Phi}^{\dagger} \mathbf{y}+\color{blue}{\left(\mathbf{I}-\mathbf{\Phi}^{\dagger} \mathbf{\Phi}\right) \mathbf{x}}\color{black}. 
\label{eq13}
\end{equation}}In HSI reconstruction, the range-space component $\mathbf{\Phi}^{\dagger} \mathbf{y}$ can actually be calculated from $\mathbf{y}$, while the null-space component $\left[\color{blue}{\left(\mathbf{I}-\mathbf{\Phi}^{\dagger} \mathbf{\Phi}\right) \mathbf{x}}\color{black}\right]$ can be refined and estimated by the network. To be noted, the combination of $\mathbf{\Phi}^{\dagger} \mathbf{y}$ and any solution $\mathbf{s}$ for the null-space component projected by $\left(\mathbf{I}-\mathbf{\Phi}^{\dagger} \mathbf{\Phi}\right)$ strictly enjoys the exact data consistency, \textit{i.e.} $\forall \mathbf{s}$, $\mathbf{\Phi}\left[\mathbf{\Phi}^{\dagger} \mathbf{y} + \color{blue}{\left(\mathbf{I}-\mathbf{\Phi}^{\dagger} \mathbf{\Phi}\right) \mathbf{s}}\color{black}\right]{ \equiv \mathbf{y}}$. Thus, our motivation is to estimate the null-space component $\left[\color{blue}{\left(\mathbf{I}-\mathbf{\Phi}^{\dagger} \mathbf{\Phi}\right) \mathbf{x}}\color{black}\right]$ only and remain the clean range-space component unchanged to alleviate the recovery difficulty of networks. Furthermore, we unfold the null-space learning in Eq.~\ref{eq13} into a deep network and resort to the proximal mapping to refine the null-space correlated component iteratively as follows.
{\setlength\abovedisplayskip{0.1cm}
\setlength\belowdisplayskip{0.1cm}
\begin{equation}
    \begin{aligned}
    \mathbf{z}^{(k)} &= \mathbf{\Phi}^{\dagger} \mathbf{y} + \color{blue}{\left(\mathbf{I}-\mathbf{\Phi}^{\dagger} \mathbf{\Phi}\right) \mathbf{x}^{(k-1)}} \\
    &= \mathbf{x}^{(k-1)} + \mathbf{\Phi}^{\dagger}(\mathbf{y}-\mathbf{\Phi}\mathbf{x}^{(k-1)}),
    \end{aligned}
    \label{eq15}
\end{equation}}
\begin{figure}[t!]
	\centering	\includegraphics[width=1.0\linewidth]{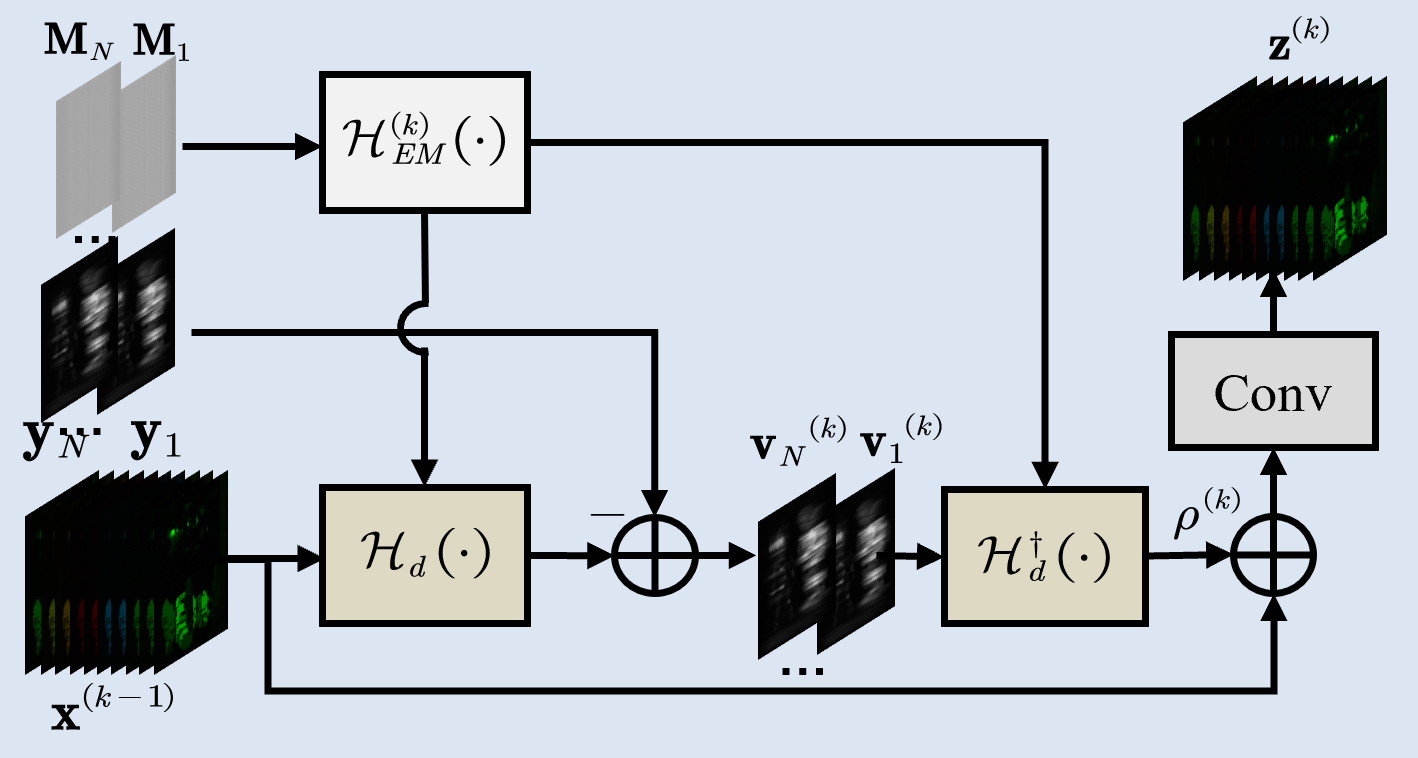}
	\caption{Details of the proposed RNDM. It retains the range-space component and recovers the null-space component of HSIs.}
	\label{detail1}
	\vspace{-15pt}
\end{figure}
{\setlength\abovedisplayskip{0.1cm}
\setlength\belowdisplayskip{0.1cm}
\begin{equation}
\mathbf{x}^{(k)}=\operatorname{prox}_{\lambda \psi}(\mathbf{z}^{(k)}),
	\label{eq16}
\end{equation}}where $\mathbf{z}^{(k)}$ and $\mathbf{x}^{(k)}$ respectively denote the result of RND and proximal mapping in the $k^{th}$ iteration. Inspired by Eq.~\ref{eq15} and Eq.~\ref{eq16}, the range-null decomposition module (RNDM) and spatial-spectral fusion module (SSFM) are proposed as follows for accurate HSI reconstruction.

\begin{table*}[]
	\renewcommand{\arraystretch}{1.}
	\centering
	\caption{Comparison results of the proposed RND-HRNet and state-of-the-art HSI reconstruction methods on the single-shot reconstruction. The best and second-best scores are \textbf{highlighted} and  \underline{underlined}.}
	\centering
	\label{table2}
	\resizebox{\textwidth}{!}{
	\begin{tabular}{c|cccccccc|cc}
	\toprule[1.5pt]
    \rowcolor[HTML]{F2F2F2} 
    \cellcolor[HTML]{F2F2F2} &\textbf{GAP-TV \cite{yuan2016generalized}} & \textbf{TSA-Net \cite{meng2020end}} & \textbf{DGSMP \cite{huang2021deep}} & \textbf{SRN \cite{wang2021new}} & \textbf{HDNet \cite{hu2022hdnet}} & \textbf{MST \cite{cai2022mask}} & \textbf{MST++ \cite{cai2022mst++}} & \textbf{CST \cite{cai2022coarse}} & \textbf{RND-HRNet} & \textbf{RND-HRNet}         \\
    \rowcolor[HTML]{F2F2F2} 
    \multirow{-2}{*}{\cellcolor[HTML]{F2F2F2}\textbf{Testing set}} & ICIP 2016 & ECCV 2020  & CVPR 2021  & Arxiv 2021 & CVPR 2022 & CVPR 2022 & CVPRW 2022 & ECCV 2022 & 2 phases & 10 phases            \\ \midrule[0.75pt]
		Scene01  &26.82 / 0.754 &32.03 / 0.892 &33.26 / 0.915 &34.85 / 0.937 &35.14 / 0.935 &35.40 / 0.941 &35.80 / 0.943 &35.16 / 0.938 &\underline{36.15} / \underline{0.948} &\textbf{37.29} / \textbf{0.959}\\
		Scene02  &22.89 / 0.610 &31.00 / 0.858 &32.09 / 0.898 &35.11 / 0.935 &35.67 / 0.940 &35.87 / 0.944 &36.24 / 0.947 &35.60 / 0.942 &\underline{37.34} / \underline{0.956} &\textbf{40.07} / \textbf{0.974}\\
		Scene03  &26.31 / 0.802 &32.25 / 0.915 &33.06 / 0.925 &35.89 / 0.949 &36.03 / 0.943 &36.51 / 0.953 &37.39 / 0.957 &36.57 / 0.953 &\underline{38.47} / \underline{0.963} &\textbf{41.48} / \textbf{0.972}\\
		Scene04  &30.65 / 0.852 &39.19 / 0.953 &40.54 / 0.964 &42.12 / 0.975 &42.30 / 0.969 &42.27 / 0.973 &43.85 / 0.973 &42.29 / 0.972 &\underline{43.95} / \underline{0.975} &\textbf{45.59} / \textbf{0.985}\\
		Scene05  &23.64 / 0.703 &29.39 / 0.884 &28.86 / 0.882 &32.53 / 0.944 &32.69 / 0.946 &32.77 / 0.947 &33.41 / 0.952 &32.82 / 0.948 &\underline{34.57} / \underline{0.959} &\textbf{36.08} / \textbf{0.971}\\
		Scene06  &21.85 / 0.663 &31.44 / 0.908 &33.08 / 0.937 &34.59 / 0.955 &34.46 / 0.952 &34.80 / 0.955 &35.43 / 0.957 &35.15 / 0.956 &\underline{35.82} / \underline{0.961} &\textbf{37.34} / \textbf{0.972}\\
		Scene07  &23.76 / 0.688 &30.32 / 0.878 &30.74 / 0.886 &33.52 / 0.924 &33.67 / 0.926 &33.66 / 0.925 &34.35 / 0.934 &33.85 / 0.927 &\underline{35.37} / \underline{0.943} &\textbf{37.27} / \textbf{0.960}\\
		Scene08  &21.98 / 0.655 &29.35 / 0.888 &31.55 / 0.923 &32.63 / 0.947 &32.48 / 0.941 &32.67 / 0.948 &33.71 / 0.953 &33.52 / 0.952 &\underline{33.95} / \underline{0.957} &\textbf{35.55} / \textbf{0.970}\\
		Scene09  &22.63 / 0.682 &30.01 / 0.890 &31.66 / 0.911 &35.04 / 0.944 &34.89 / 0.942 &35.39 / 0.949 &36.67 / 0.953 &35.28 / 0.946 &\underline{37.57} / \underline{0.961} &\textbf{39.99} / \textbf{0.972}\\
		Scene10  &23.10 / 0.584 &29.59 / 0.874 &31.44 / 0.925 &31.99 / 0.938 &32.38 / 0.937 &32.50 / 0.941 &33.38 / 0.945 &32.84 / 0.940 &\underline{33.46} / \underline{0.945} &\textbf{34.69} / \textbf{0.960}\\
		Average  &24.36 / 0.669 &31.46 / 0.894 &32.63 / 0.917 &34.82 / 0.945 &34.97 / 0.943 &35.18 / 0.948 &36.02 / 0.951 &35.31 / 0.947 &\underline{36.66} / \underline{0.957} &\textbf{38.54} / \textbf{0.969}      \\ \bottomrule[1.5pt]
	\end{tabular}}
	\vspace{-10pt}
\end{table*}

\begin{figure*}[!]
	\centering
	\includegraphics[width=1\linewidth]{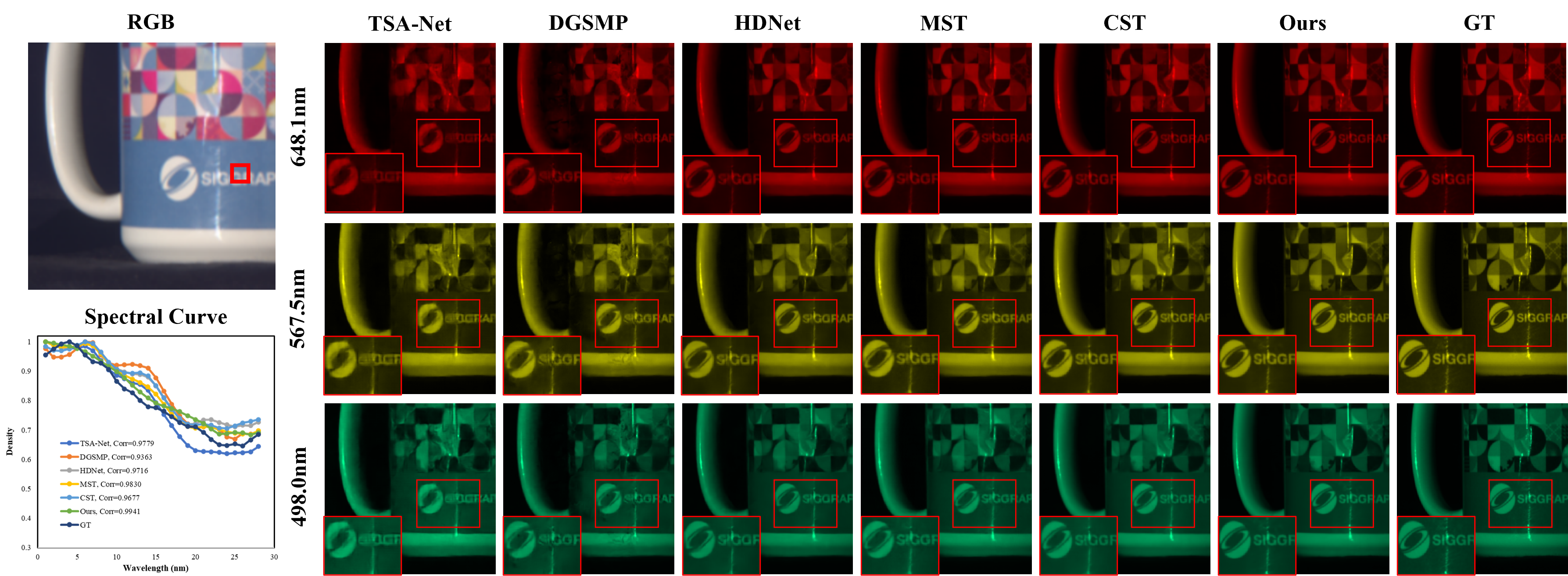}
	\vspace{-10pt}
	\caption{The visual comparison of the proposed RND-HRNet and other SOTA methods on the single-shot reconstruction.}
	\label{compare2}
 \vspace{-10pt}
\end{figure*}
\begin{figure}[t!]
	\centering
	\includegraphics[width=1\linewidth]{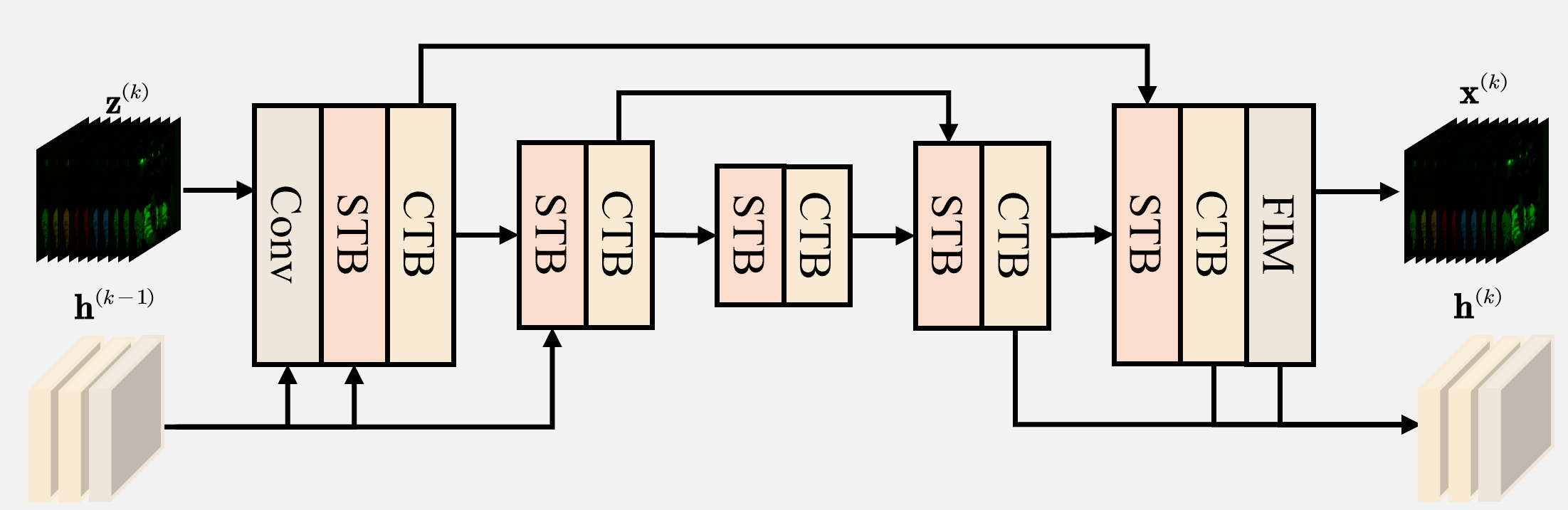}
	\caption{Details of the proposed SSFM. It extracts the spectral-spatial features with high-throughput transmission $\mathbf{h}^{(k-1)}$.}
	\label{detail2}
	\vspace{-15pt}
\end{figure}
\subsubsection{Range-Null Space Decomposition Module}
Inspired by Eq.~\ref{eq15}, our RNDM aims to retain the range space and refine the null space of the HSIs iteratively (Fig.~\ref{detail1}). Given the snapshots $\{\mathbf{y}_i\}_{i=1}^{N}$ and masks $\{\mathbf{M}_i\}_{i=1}^{N}$, RNDM yields the intermediate result $\mathbf{z}^{(k)}$ flexibly. 
{\setlength\abovedisplayskip{0.1cm}
\setlength\belowdisplayskip{0.1cm}
\begin{equation}
    \mathbf{z}^{(k)}= \operatorname{RNDM}^{(k)}(\mathbf{x}^{(k-1)}, \mathbf{y}_1,\ldots, \mathbf{y}_N,\mathbf{M}_1, \ldots, \mathbf{M}_N).
\end{equation}}

Due to the cumulative errors caused by equipment deviation, noise corruption and alignment of the continuous spectrum in real scenes, using physical masks $\{\mathbf{M}_i\}_{i=1}^{N}$ directly may be imprecise. Therefore, a deep enhanced module $\mathcal{H}_{EM}^{(k)}(\cdot)$ is introduced to produce two enhanced mask representations $\mathbf{F}_i^{(k)}$$\in$$\mathbb{R}^{H\times W \times C}$ and $\mathbf{E}_i^{(k)}$$\in$$ \mathbb{R}^{H\times W \times C}$ from the physical mask ${\mathbf{M}}_i$ $\in$$ \mathbb{R}^{H\times W}$, which corrects the bias between the degradation process and the physical mask via an attention mechanism \cite{cai2022mask} as follows.
{\setlength\abovedisplayskip{0.1cm}
\setlength\belowdisplayskip{0.1cm}
\begin{equation}
     \mathbf{F}_i^{(k)}, \mathbf{E}_i^{(k)} = \mathcal{H}_{EM}^{(k)}(\mathbf{M}_{i}).
    \end{equation}}

Furthermore, the crucial step of RNDM is to solve the degradation operator $\mathcal{H}_d(\cdot, \cdot)$ and its pseudo-inverse operator $\mathcal{H}_d^{\dagger}(\cdot, \cdot)$. Specifically, $\mathcal{H}_d(\cdot, \cdot)$ simulates the process of mask modulation, dispersion, and compression in Sec.~\ref{sec:cassi} with the current state $\mathbf{x}^{(k-1)}$ and enhanced mask representations $\mathbf{F}_i^{(k)}$. $\mathcal{H}_d^{\dagger}(\cdot, \cdot)$ is designed to provide an initialization from 2D signals to 3D cubes with the enhanced representation $\mathbf{E}_i^{(k)}$. \textbf{More details about these two operators are presented in \textcolor{red}{SM}}. To dynamically balance the contribution of range- and null-space signals, the learnable parameter $\rho^{(k)}$ is incorporated into the optimization process. Hence, we can do  $\mathcal{R}$$-$$\mathcal{N}$ decomposition on each $\mathbf{y}_i$ to get the intermediate results $\{\mathbf{z}_i^{(k)}\}_{i=1}^{N}$. Then, an $1$$\times$$1$ convolution is utilized to adaptively merge $\{\mathbf{z}_i^{(k)}\}_{i=1}^{N}$ into $\mathbf{z}^{(k)}$ as follows, which is fed to the subsequent proximal mapping module.  
{\setlength\abovedisplayskip{0.1cm}
\setlength\belowdisplayskip{0.1cm}
\begin{equation}
\begin{aligned}
    \mathbf{v}_i^{(k)} &= \mathbf{y}_i-\mathcal{H}_{d}(\mathbf{x}_{i}^{(k-1)}, \mathbf{F}_{i}^{(k)}), \\
		\mathbf{z}_{i}^{(k)} &=\mathbf{x}_{i}^{(k-1)}+\rho^{(k)}\mathcal{H}_d^{\dagger}(\mathbf{v}_i^{(k)}, \mathbf{E}_i^{(k)}), \\
      \mathbf{z}^{(k)} &=\operatorname{Conv}^{(k)}(\left[\mathbf{z}_1^{(k)}, ..., \mathbf{z}_N^{(k)}\right]),
		\label{eq19}
\end{aligned}
\end{equation}}where $\mathbf{v}_i^{(k)}$ and $[\cdot]$ respectively denote the auxiliary variable and the channel-wise concatenation operation. The measurements compressed by various coded patterns reflect different HSI contents and fuse complementary information.

\subsubsection{Spectral-Spatial Fusion Module}
To implement Eq.~\ref{eq16} with deep networks, SSFM is adopted to refine the fused intermediate result $\mathbf{z}^{(k)}$ and yield the reconstruction result $\mathbf{x}^{(k)}$ (Fig.~\ref{detail2}). Given that HSI representations are spatially sparse and spectrally correlated, capturing spatial interactions and spectral correlation are just as important. Inspired by the previous works \cite{cai2022mask, zhang2022practical}, the channel-wise and spatial-wise transformer blocks are plugged into the U-shaped architecture. The spatial-wise transformer block (STB) fuses the swin transformer block and residual convolution blocks to integrate local and non-local information. The channel-wise transformer block (CTB) encodes each feature frame into a token and explores spectral self-attention. To avoid information loss and model degradation, a feature interaction model (FIM) \cite{zhang2022herosnet} is incorporated into the SSFM to interact with the spatial-spectral features in other phases. Furthermore, since feature maps at each
scale of the U-shaped architecture also have well-preserved spatial information, the multi-scale features in other layers are also utilized for feature fusion. Noted that the concatenation of the spatial-spectral features and multi-scale features in the $k^{th}$ phase are denoted by $\mathbf{h}^{(k)}$. The cascaded features in the previous phase are conducive to the reconstruction of the current phase. \textbf{More details are presented in \textcolor{red}{SM}}.
Finally, our SSFM is summarized as:
{
\setlength\abovedisplayskip{0.1cm}
\setlength\belowdisplayskip{0.1cm}
\begin{equation}
\mathbf{h}^{(k)}, \mathbf{x}^{(k)}= \operatorname{SSFM}^{(k)} (\mathbf{z}^{(k)}, \mathbf{h}^{(k-1)}).
	\label{eq20}
\end{equation}}

\begin{table*}[]
	\renewcommand{\arraystretch}{1.}
	\centering
	\caption{Comparison results of the proposed method and state-of-the-art HSI reconstruction methods on the two-shot reconstruction. The best and second-best scores are \textbf{highlighted} and  \underline{underlined}. \textcolor{blue}{Here, $\mathcal{H}_R(\cdot)$ in PCA-CASSI and RND-HRNet both include 2 phases.}}
	\centering
	\label{table1}
	\resizebox{\textwidth}{!}{
	\begin{tabular}{c|cccccccc|cc}
        \toprule[1.5pt]
    \rowcolor[HTML]{F2F2F2} 
    \cellcolor[HTML]{F2F2F2} &\textbf{GAP-TV \cite{yuan2016generalized}} &\textbf{ADMM-TV \cite{yuan2016generalized}} &\textbf{TSA-Net \cite{meng2020end}} & \textbf{SRN \cite{wang2021new}} &\textbf{HDNet \cite{hu2022hdnet}} &\textbf{MST \cite{cai2022mask}} &\textbf{MST++ \cite{cai2022mst++}} &\textbf{CST \cite{cai2022coarse}} &\textbf{RND-HRNet} &\textbf{PCA-CASSI}        \\
    \rowcolor[HTML]{F2F2F2} 
    \multirow{-2}{*}{\cellcolor[HTML]{F2F2F2}\textbf{Testing set}} 	&ICIP 2016 &ICIP 2016 &ECCV 2020 &Arxiv 2021 &CVPR 2022 &CVPR 2022 &CVPRW 2022 &ECCV 2022 &Ours &Ours            \\
        \midrule[0.75pt]
		Scene01  &27.62 / 0.739 &27.85 / 0.768 &33.62/0.912 & 34.79 / 0.930 &35.20 / 0.941 &35.50 / 0.942 &35.89 / 0.947 & 35.57 / 0.939 &\underline{37.49} / \underline{0.961} & \textbf{39.49} / \textbf{0.970}\\
		Scene02  &25.92 / 0.665 &25.65 / 0.684 &32.51/0.893 & 35.09 / 0.920 &36.00 / 0.945 &36.00 / 0.939 &36.70 / 0.946 & 35.99 / 0.940 &\underline{39.78} / \underline{0.969} & \textbf{43.52} / \textbf{0.984}\\
		Scene03  &23.65 / 0.762 &23.84 / 0.791 &33.89/0.932 & 34.50 / 0.930 &35.01 / 0.936 &36.38 / 0.943 &36.44 / 0.947 & 35.94 / 0.943 &\underline{39.01} / \underline{0.957} & \textbf{40.91} / \textbf{0.968}\\
		Scene04  &34.20 / 0.872 &33.56 / 0.886 &40.40/0.964 & 40.94 / 0.955 &41.19 / 0.968 &43.16 / 0.975 &43.56 / 0.975 & 42.32 / 0.965 &\underline{44.49} / \underline{0.980} & \textbf{47.44} / \textbf{0.988}\\
		Scene05  &23.90 / 0.708 &23.94 / 0.732 &31.05/0.917 & 32.21 / 0.925 &32.76 / 0.946 &33.31 / 0.946 &33.18 / 0.948 & 33.25 / 0.945 &\underline{35.62} / \underline{0.965} & \textbf{38.45} / \textbf{0.978}\\
		Scene06  &23.97 / 0.670 &23.85 / 0.698 &33.04/0.930 & 35.63 / 0.938 &36.10 / 0.958 &35.91 / 0.954 &36.52 / 0.960 & 36.36 / 0.955 &\underline{38.47} / \underline{0.970} & \textbf{40.15} / \textbf{0.980}\\
		Scene07  &23.46 / 0.682 &23.60 / 0.712 &32.06/0.902 & 33.63 / 0.919 &33.44 / 0.917 &33.70 / 0.916 &33.96 / 0.918 & 34.02 / 0.921 &\underline{36.27} / \underline{0.945} & \textbf{38.72} / \textbf{0.963}\\
		Scene08  &23.68 / 0.654 &23.93 / 0.695 &31.37/0.919 & 34.45 / 0.933 &34.96 / 0.958 &35.07 / 0.951 &35.58 / 0.961 & 35.62 / 0.954 &\underline{37.58} / \underline{0.971} & \textbf{39.48} / \textbf{0.981}\\
		Scene09  &25.18 / 0.708 &25.04 / 0.743 &32.27/0.913 & 34.27 / 0.930 &34.42 / 0.937 &36.15 / 0.944 &35.89 / 0.945 & 35.52 / 0.942 &\underline{38.69} / \underline{0.963} & \textbf{40.35} / \textbf{0.973}\\
		Scene10  &24.22 / 0.589 &24.54 / 0.603 &30.42/0.887 & 33.82 / 0.942 &33.95 / 0.958 &34.50 / 0.954 &34.67 / 0.958 & 34.83 / 0.956 &\underline{36.54} / \underline{0.971} & \textbf{38.78} / \textbf{0.981}\\
		Average  &25.58 / 0.705 &25.58 / 0.731 &33.06/0.917 & 34.93 / 0.932 &35.30 / 0.947 &35.97 / 0.946 &36.24 / 0.951 & 35.94 / 0.946 &\underline{38.39} / \underline{0.965} & \textbf{40.73} / \textbf{0.977}        \\ \bottomrule[1.5pt]
	\end{tabular}}
	\vspace{-5pt}
\end{table*}
\begin{figure*}[!]
	\centering
	\includegraphics[width=1\linewidth]{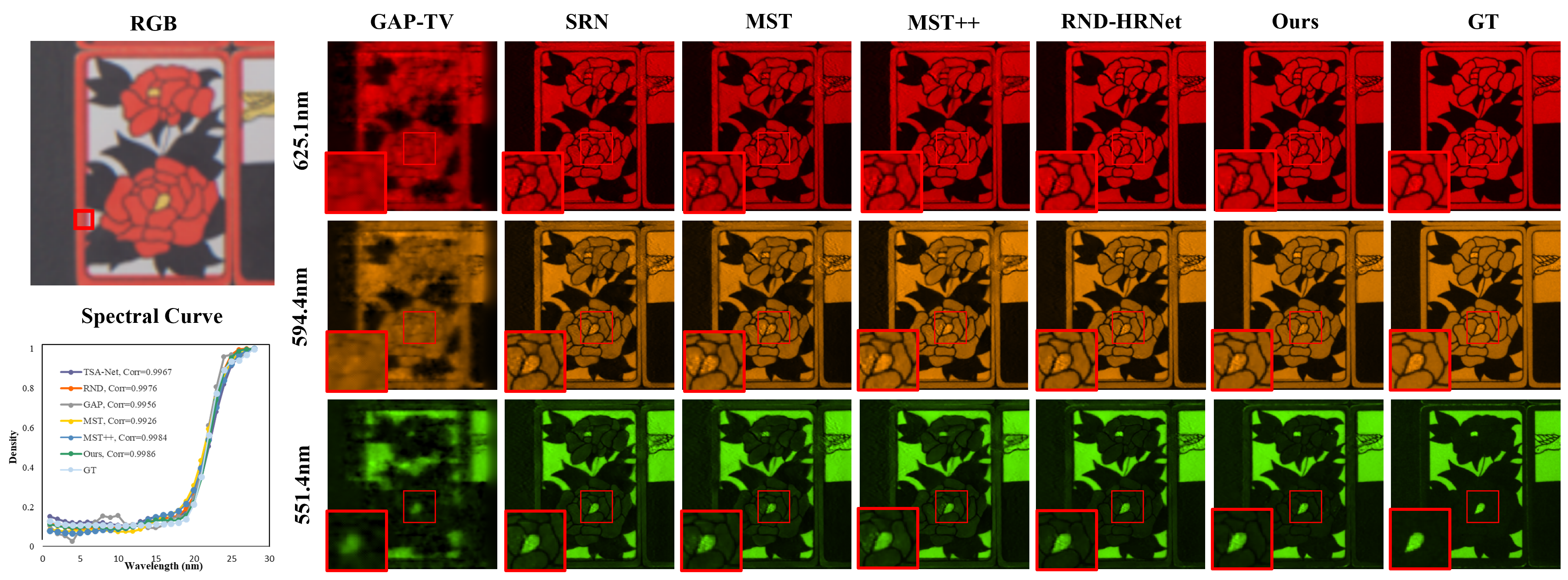}
	\caption{The visual comparison of the proposed method and other SOTA methods on the multiple-shot reconstruction.}
	\label{compare1}
	\vspace{-10pt}
\end{figure*}

\subsection{Network Training and Implementation}
Without bells and whistles, the proposed reconstruction network $\mathcal{H}_R(\cdot)$ and mask predictors $\{\mathcal{H}_M^{(i)}(\cdot)\}_{i=1}^{N}$ are jointly optimized via a common MSE loss. Specifically, given the training data $\{\mathbf{x}_{i}\}_{i=1}^{N_{d}}$, the loss function is defined:
{\setlength\abovedisplayskip{0.1cm}
\setlength\belowdisplayskip{0.1cm}
\begin{equation}
	\mathcal{L}(\mathbf{\Theta})=\frac{1}{HWCN_{d}} \sum_{i=1}^{N_{d}}\|\hat{\mathbf{x}}_i -\mathbf{x}_{i}\|_{2}^{2},
	\label{eq21}
\end{equation}}where $\hat{\mathbf{x}}_i$ and $N_{d}$ are the recovered result and training sample number. $\mathbf{\Theta}$$=$$\left[\bigcup_{i \in\{1, \cdots, N\}} \mathcal{H}_M^{(i)}\right] \cup \mathcal{H}_R$ denotes the set of all learnable parameters with $\mathcal{H}_M^{(i)}$$=$$\{\overline{\mathbf{M}_i}, \eta^{(i)}, \mathcal{H}_M^{\prime(i)}\}$, $\mathcal{H}_R$$=$$\{\operatorname{RNDM}^{(k)}, \operatorname{SSFM}^{(k)}\}_{k=1}^K$, where $\operatorname{RNDM}^{(k)}$$=$$\{\rho^{(k)}, \mathcal{H}_{EM}^{(k)}, \text{Conv}^{(k)}\}$. All the parameters are indiscriminately learned in an end-to-end manner. The Adam \cite{kingma2014adam} optimizer is employed for the training of 200 epochs. The learning rate is initialized to $4$$\times$$10^{-4}$ and scheduled to $1$$\times$$10^{-6}$ using cosine annealing.

\section{Experimental Results}
\label{sec:expr}
\subsection{Experimental Settings}
In this paper, the effectiveness of the proposed method has been verified on both simulation datasets and the real dataset. Following \cite{meng2020end}, the simulation experiments are conducted on the public HSI datasets CAVE \cite{yasuma2010generalized} and KAIST \cite{choi2017high} with the size $256$$\times$$256$$\times$$28$. Meanwhile, 5 real compressive measurements \cite{huang2021deep} with the size of $660$$\times$$714$ are used for testing. The metrics of PSNR and SSIM \cite{wang2004image} are employed to evaluate the reconstruction quality.

\subsection{Simulation Results}
To demonstrate the effectiveness of the proposed PCA-CASSI and RND-HRNet, we compare our methods with other state-of-the-art (SOTA) methods, including model-based methods \cite{yuan2016generalized}, classical CNN-based methods \cite{wang2021new, meng2020end, huang2021deep, hu2022hdnet} and recent transformer networks \cite{cai2022mask,cai2022coarse, cai2022mst++}. 

\noindent \textbf{Single-shot Reconstruction:} 
Following the settings of previous works \cite{meng2020end, huang2021deep}, we conduct the single-shot reconstruction on the KAIST dataset \cite{choi2017high}. It can be clearly seen that the proposed RND-HRNet with 10 phases yields 38.54 dB on PSNR and 0.969 on SSIM, which significantly outperforms the most recent works \cite{cai2022mask, cai2022coarse, cai2022mst++} \textbf{over 2dB}. To be noted, even with only 2 phases, the proposed RND-HRNet can also achieve the current best performance. As exhibited in Fig.~\ref{compare2}, our method can recover sharper edges and more realistic details, which indicates that the proposed RNDM and SSFM can accurately maintain data consistency and effectively exploit the non-local correlation. 

\begin{table}[]
\centering
\caption{Evaluation of computational complexity on multiple- and single-shot reconstructions.``Ph'' denotes the number of Phases.}
\resizebox{!}{1.7cm}{
\begin{tabular}{c|c|c|c|c|c}
\toprule[1.5pt]
\rowcolor[HTML]{F2F2F2} Method & Shot & Testing time &Params. &PSNR &SSIM \\ \midrule[0.75pt]
HDNet \cite{hu2022hdnet} & $\#1$  & 116.92ms  &2.37M &34.97dB &0.943\\
MST \cite{cai2022mask}   & $\#1$  & 176.36ms  &2.03M &35.18dB &0.948\\
CST \cite{cai2022coarse} & $\#1$  & 153.18ms  &1.36M &35.31dB &0.947\\
RND-HRNet-2Ph      & $\#1$  & 189.13ms  &1.82M &36.66dB &0.957 \\
RND-HRNet-10Ph     & $\#1$  & 531.37ms  &9.48M &38.54dB &0.969      \\
HDNet \cite{hu2022hdnet} & $\#2$  & 117.45ms  &2.37M &35.30dB &0.947\\
MST \cite{cai2022mask}   & $\#2$  & 179.25ms  &2.03M &35.97dB &0.946\\
CST \cite{cai2022coarse}   & $\#2$  & 159.21ms  &1.36M &35.94dB &0.946\\
RND-HRNet-2Ph      & $\#2$  & 199.21ms  &1.82M &38.39dB &0.965       \\
PCA-CASSI-2Ph      & $\#2$  & 249.45ms  &1.84M &40.73dB &0.977\\
\bottomrule[1.5pt]
\end{tabular}}
\label{complexity}
\vspace{-15pt}
\end{table}

\noindent \textbf{Multiple-shot Reconstruction:} In this section, we extend the RND-HRNet to multiple-shot with the proposed progressive content-aware sampling. For PCA-CASSI, two optimized masks are utilized to capture HSIs respectively. For other methods, we adopt two fixed real masks, namely $\mathbf{M}$ \cite{meng2020end} and $1$$-$$\mathbf{M}$, to realize the two-shot imaging. To adapt existing methods to multiple-shot reconstruction, we utilize a few $1\times1$ convolutions to fuse two measurements and masks in learning-based methods (SRN, HDNet, MST, CST). Besides, for model-based methods (GAP-TV, ADMM-TV), we jointly solve these two sub-optimization problems. 

\begin{figure}[t!]
	\centering
	\includegraphics[width=1\linewidth]{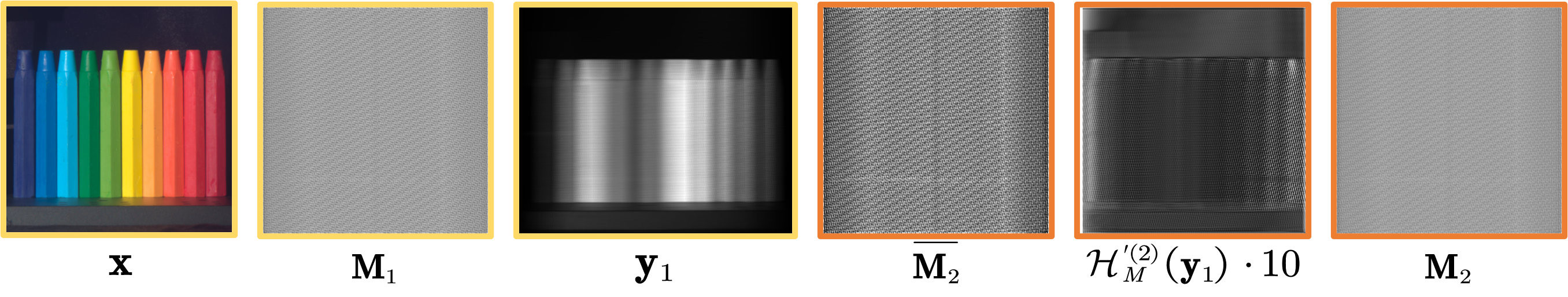}
	\vspace{-15pt}
	\caption{Visualization results of the optimized $\mathbf{M}_1$ and $\mathbf{M}_2$, where $\mathbf{M}_2=\overline{\mathbf{M}_2}+\eta^{(2)} \cdot \mathcal{H}_M^{\prime(2)}(\mathbf{y}_1)$. For clearer presentation, all the element values of $\mathcal{H}_M^{\prime(2)}(\mathbf{y}_1)$ are magnified by 10.}
	\label{mask}
 \vspace{-10pt}
\end{figure}

As shown in Tab.~\ref{table1}, all reconstruction methods have achieved corresponding improvements compared to single-shot results listed in Tab.~\ref{table2}. However, since the fusion method does not fully utilize the complementary information of multiple masks, some methods do not reach their limit of performance. Owing to the proposed RND-inspired fusion mechanism, RND-HRNet with 2 phases achieves 38.39dB on PSNR and 0.965 on SSIM, which surpasses 1.73 dB than the single-shot case. Noted that even utilizing the exact same system and masks, the proposed RND-HRNet also outperforms the SOTA method MST++ \cite{cai2022mst++} by 2.15 dB on PSNR. Considering the trade-off between computational complexity and performance, RND-HRNet with 10 phases is not conducted on the multiple-shot case. Furthermore, utilizing the content-aware optimized masks, the proposed PCA-CASSI achieves very impressive results, \textit{i.e.} 40.73dB on PSNR and 0.977 on SSIM. Compared with two recent SOTA methods CST \cite{cai2022coarse} and MST \cite{cai2022mask}, our method surpasses them by 4.79dB and 4.76dB on PSNR, which verifies the effect of the proposed imaging framework. As depicted in Fig.~\ref{compare1}, the HSIs produced by our methods have clearer spatial details and more accurate spectral consistency. Meanwhile, Fig.~\ref{mask} presents the content-aware optimized masks. An interesting finding is that the optimized mask $\mathbf{M}_2$ contains a shared pattern $\overline{\mathbf{M}_2}$ and veils some content-aware textures $\mathcal{H}_M^{\prime(2)}(\mathbf{y}_1)$.

\noindent \textbf{Computational complexity:} As listed in Tab.~\ref{complexity}, the proposed RND-HRNet achieves greater performance than that of CST and MST with the same level of average testing time and even less parameters, which verifies the effect of the proposed RND-HRNet. With the mask predictor, the proposed method with 2 phases can obtain 2.34dB gains on PSNR with only 0.02M increase in parameters, proving the practicality of the progressive content-aware sampling.

\begin{figure}[t!]
	\centering
	\includegraphics[width=1\linewidth]{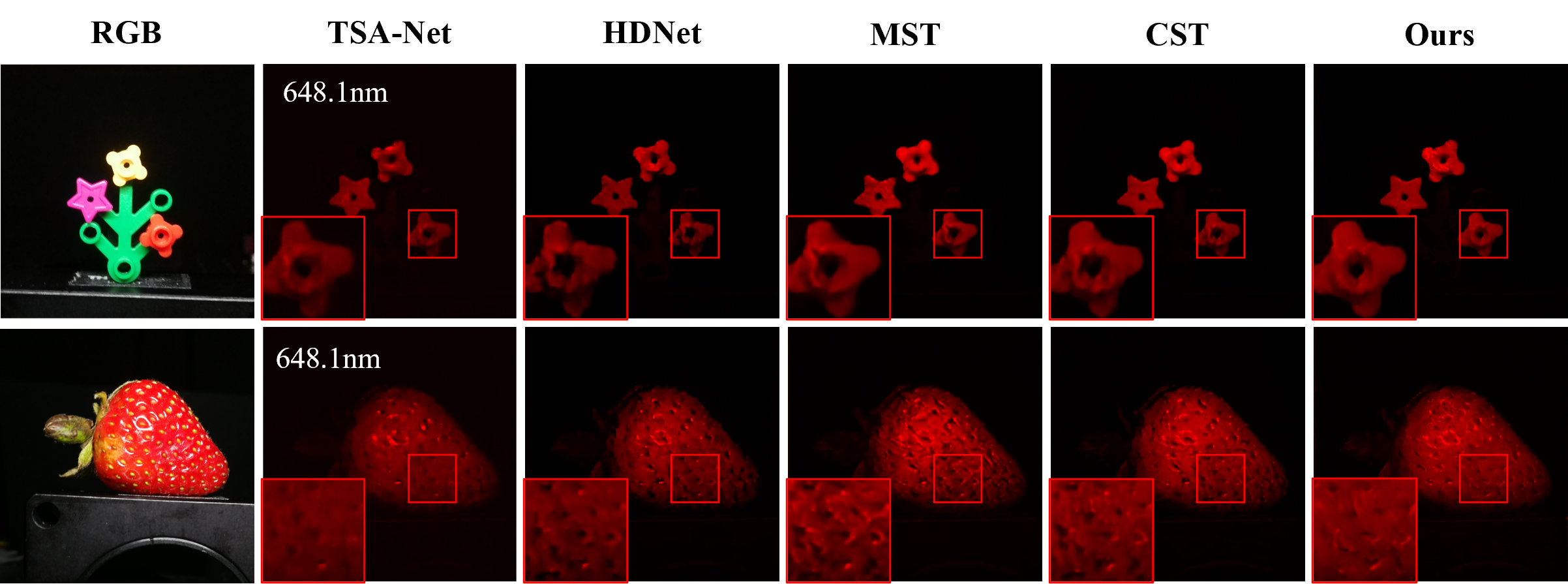}
	\vspace{-10pt}
	\caption{Real data results of the proposed RND-HRNet and other SOTA methods on the single-shot reconstruction.}
	\label{real}
 \vspace{-10pt}
\end{figure}

\subsection{Real Data Results} 
To objectively evaluate the proposed method on real data, RND-HRNet is tested in 5 real measurements \cite{huang2021deep}. To simulate the real imaging situations, 11-bit shot noise is injected into the training data \cite{meng2020end}. As shown in Fig.~\ref{real}, our results can reconstruct clearer HSI contents and detailed textures with fewer artifacts and noise, especially in the cropped regions such as flowers and strawberries, suggesting the generalization ability and robustness of our method.

\subsection{Ablation Study}
To evaluate the contribution of the proposed PCA-CASSI, we deploy the CASSI system on different mask combinations. For convenience, the reconstruction network is an one-phase RND-HRNet. Tab.~\ref{table3} indicates the PSNR and SSIM on different mask combinations. 

\noindent \textbf{Multiple-shot vs. Single-shot:} Comparing case 1 and case 4 or case 2 and case 5, HSI reconstruction with two fixed or optimized masks surpasses one-shot reconstruction by 0.69dB or 2.86dB on PSNR, suggesting that multiple complementary coded apertures capture richer information.

\noindent \textbf{Optimized vs. Fixed:} As shown in Tab.~\ref{table3}, the more optimized masks are used, the better the reconstruction performance is, indicating that more optimized masks better remove redundancy and retain useful information (case 2,3,5). 

\noindent \textbf{Content-aware vs. Shared:} Comparing case 5 and case 6, we find that although randomly optimized masks are conducive to information acquisition, the same coded apertures are shared and non-adaptive for all HSI scenes, which inevitably leads to the loss of spectral anisotropy. With the proposed progressive content-aware strategy, HSI reconstruction with content-aware optimized masks can surpass the results with shared optimized masks by 0.43dB, which proves the effect of our mask optimization algorithm.

\begin{table}[]
\caption{Evaluation of different mask combinations. \textcolor{blue}{Here, the reconstruction network includes 1 phase for simplicity.} More results of the $N$-shot cases ($N \geq 3$) are presented in our \textcolor{red}{SM}.}
\centering
\resizebox{!}{1.5cm}{
\begin{tabular}{c|cc|c|c|c}
\toprule[1.5pt]
\rowcolor[HTML]{F2F2F2} 
\multicolumn{1}{c|}{\cellcolor[HTML]{F2F2F2}}                       & \multicolumn{2}{c|}{\cellcolor[HTML]{F2F2F2}Mask   Type}                                                              & \multicolumn{1}{c|}{\cellcolor[HTML]{F2F2F2}}                                & \multicolumn{1}{c|}{\cellcolor[HTML]{F2F2F2}}                       & \multicolumn{1}{c}{\cellcolor[HTML]{F2F2F2}}                       \\ \cline{2-3}
\rowcolor[HTML]{F2F2F2} 
\multicolumn{1}{c|}{\multirow{-2}{*}{\cellcolor[HTML]{F2F2F2}Case}} & \multicolumn{1}{c|}{\cellcolor[HTML]{F2F2F2}Fixed mask} & \multicolumn{1}{c|}{\cellcolor[HTML]{F2F2F2}Optimized mask} & \multicolumn{1}{c|}{\multirow{-2}{*}{\cellcolor[HTML]{F2F2F2}Content-aware}} & \multicolumn{1}{c|}{\multirow{-2}{*}{\cellcolor[HTML]{F2F2F2}PSNR}} & \multicolumn{1}{c}{\multirow{-2}{*}{\cellcolor[HTML]{F2F2F2}SSIM}} \\ \midrule[0.75pt]
1                     & \multicolumn{1}{c|}{1}          & 0              & $\usym{2717}$                              & 35.04                 & 0.948                 \\
2                     & \multicolumn{1}{c|}{0}          & 1              & $\usym{2717}$                              & 36.33                 & 0.957                 \\
3                     & \multicolumn{1}{c|}{1}          & 1              & $\usym{2717}$                             & 38.29                 & 0.967                 \\
4                     & \multicolumn{1}{c|}{2}          & 0              & $\usym{2717}$                              & 35.73                 & 0.953                 \\
5                     & \multicolumn{1}{c|}{0}          & 2              & $\usym{2717}$                              & 39.19                 & 0.972                 \\
6                     & \multicolumn{1}{c|}{0}          & 2              & $\usym{2713}$                              & 39.62                 & 0.974                 \\ \bottomrule[1.5pt]
\end{tabular}}
\label{table3}
\vspace{-10pt}
\end{table}

\vspace{-5pt}
\section{Conclusion}
In this paper, we propose a novel spectral snapshot compressive imaging framework, dubbed PCA-CASSI. It progressively compresses HSIs with content-aware optimized coded apertures and fuses all the snapshots for reconstruction. Inspired by the $\mathcal{R}$$-$$\mathcal{N}$ decomposition, an RND-HRNet is proposed for accurate HSI reconstruction. To improve its representation ability, a range-null space decomposition module is proposed to refine the null-space component of HSIs iteratively. Meanwhile, a spatial-spectral fusion module is introduced to explore the non-local correlation via dual transformer blocks. Extensive experiments verify that our method significantly outperforms the other SOTA methods on both multiple- and single-shot reconstruction tasks.

{\small
\bibliographystyle{ieee_fullname}
\bibliography{egbib}
}

\appendix
\onecolumn
\begin{appendices}

\begin{figure}[H]
	\centering
    \includegraphics[width=1\textwidth]{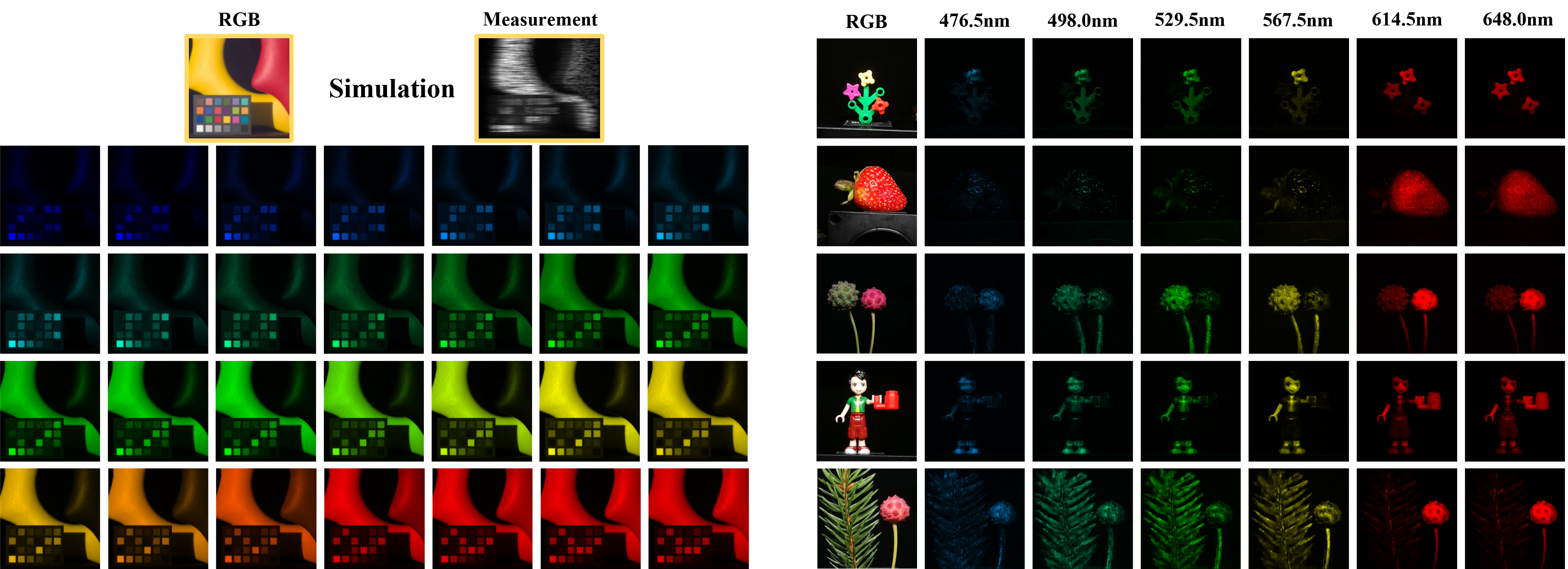}
    \caption{Reconstructed results of the proposed methods. Simulation results of all 28 bands \cite{huang2021deep} are presented in the left column and the real data results are shown in the right column.}
    \label{logo}
\end{figure}

In the supplementary materials, we demonstrate additional experimental results and implementation details as follows.
\begin{itemize}
    \setlength{\itemsep}{0.3pt}
    \setlength{\parsep}{0.3pt}
    \setlength{\parskip}{0.3pt}
    \item Implementation details of some key modules are elaborated in Sec.~\ref{sec: id}; 
    \item More visualizations of the real reconstruction results, simulation results, and content-aware masks are shown in Sec.~\ref{sec: results};
    \item Ablation studies on the proposed RND-HRNet and $N$-shot reconstruction (with $N > 2$) are conducted in Sec.~\ref{sec: aba};
    \item The limitations and broader impacts of this work are discussed and presented in Sec.~\ref{sec: limit}.
\end{itemize}

\section{Implementation Details of some Key Modules}
\label{sec: id}
\subsection{The pseudo codes of \texorpdfstring{$\mathcal{H}_d(\cdot, \cdot)$ } \textit{and} \texorpdfstring{$\mathcal{H}_d^{\dagger}(\cdot, \cdot)$ } \textit{in} the RNDM}
To further reflect the forward and inverse operation of the imaging system, the PyTorch-style pseudo-codes of $\mathcal{H}_d(\cdot, \cdot)$ and $\mathcal{H}_d^{\dagger}(\cdot, \cdot)$ in the range-nullspace decomposition module (RNDM) are presented as follows.

\lstset{ %
language=python,                
basicstyle=\small,           
numbers=left,                   
numberstyle=\tiny\color{gray},  
stepnumber=0,                   
numbersep=14pt,                  
backgroundcolor=\color{white},      
showspaces=false,               
showstringspaces=false,         
showtabs=false,                 
frame=single,                   
rulecolor=\color{black},        
tabsize=2,                      
captionpos=b,                   
breaklines=true,                
breakatwhitespace=false,        
title=\lstname,                 
keywordstyle=\color{blue},          
commentstyle=\color{green},       
stringstyle=\color{orange},         
escapeinside={\%*}{*)},            
morekeywords={*,...}               
}
\begin{lstlisting}
def H_d(x, F): # shift and sum 
    # input: x; shape: (B, C, H, W)  F; shape: (B, C, H, W)
    # output: y; shape: (B, 1, H, W+(C-1)*step)
    step = 2
    z = x * F
    y = zeros(B, 1, H, W)
    for i in range(C):
        y[:, :, :, i*step:i*step + W] += z[:, i:i+1]
    return y
    
def H_d_pinv(y, E): # rectification, cropping and sliding
    # input: y; shape: (B, 1, H, W+(C-1)*step) E; shape: (B, C, H, W)
    # output: x; shape: (B, C, H, W)
    step = 2
    alpha = zeros(B, 1, 1, W+(C-1)*step)
    for i in range(C):
        alpha[:, :, :, i*step:i*step + W] += 1
    alpha = 1 / alpha
    z = y * alpha
    x = []
    for i in range(C):
        x.append(z[:, :, :, i*step:i*step + W])
    x = Concatenate(x, dim=1) / E
    return x
\end{lstlisting}
\subsection{Details of the CTB and STB in the SSFM}
\begin{figure}[H]
	\centering
	\includegraphics[width=1\linewidth]{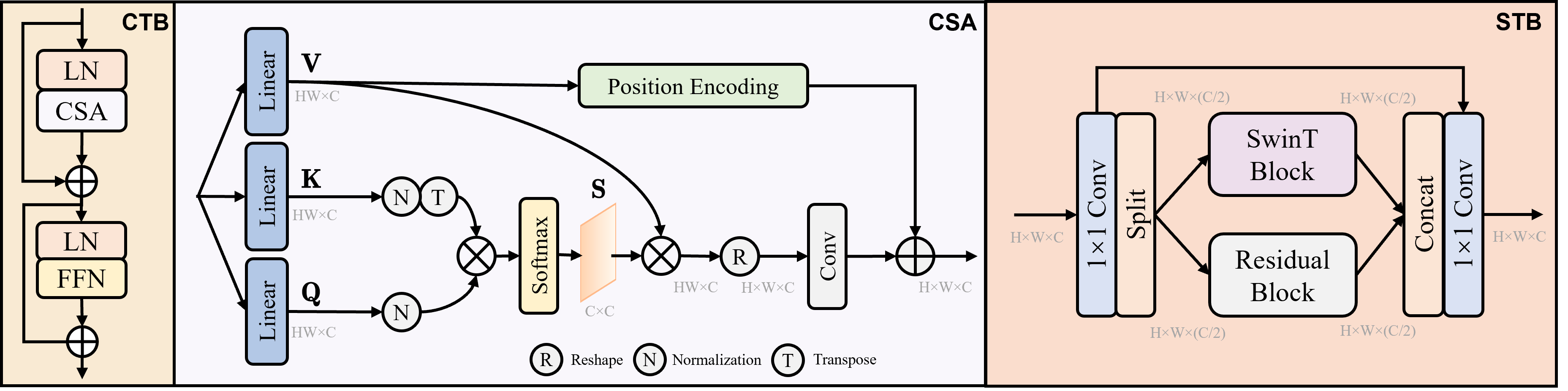}
	\vspace{-5pt}
	\caption{The details of channel-wise transformer block (CTB) and spatial-wise transformer block (STB) in the spectral-spatial fusion module (SSFM). These two blocks allow the SSFM to explore spectral and spatial correlation and fuse local and non-local information.}
	\label{detail}
\end{figure}

In this subsection, we elaborate on the details of the CTB and STB in Fig.~\ref{detail} to show how the SSFM fuses spatial and spectral correlations adaptively. Following the previous work \cite{cai2022mask}, the channel-wise transformer block (CTB) aims to extract spectral correlation, which includes two-layer normalization (LN) layers, one channel-wise self-attention (CSA) block and one feed-forward network (FFN). For convenience, we assume that the batch size is set to 1 respectively. Concretely, the CSA block encodes each feature frame $\mathbf{I}_c$ to the channel-wise tokens via three linear layers to obtain the value $\mathbf{V} \in \mathbb{R}^{HW \times C}$, key $\mathbf{K} \in \mathbb{R}^{HW \times C}$ and query $\mathbf{Q} \in \mathbb{R}^{HW \times C}$ as follows.
\begin{equation}
\mathbf{Q}=\mathbf{I}_c \mathbf{W}_{\mathbf{Q}}, \mathbf{K}=\mathbf{I}_c \mathbf{W}_{\mathbf{K}}, \mathbf{V}=\mathbf{I}_c \mathbf{W}_{\mathbf{V}},
\end{equation}
where $\mathbf{W}_{\mathbf{Q}}$, $\mathbf{W}_{\mathbf{K}}$ and $\mathbf{W}_{\mathbf{V}}$$\in$$\mathbb{R}^{C \times C}$ respectively denote the learnable parameters. Then, $\mathbf{Q}$, $\mathbf{K}$ and $\mathbf{V}$ are respectively split into $N$ heads along the channel dimension, namely $\mathbf{Q}=\left[\mathbf{Q}_1, \ldots, \mathbf{Q}_N\right]$, $\mathbf{K}=\left[\mathbf{K}_1, \ldots, \mathbf{K}_N\right]$ and $\mathbf{V}=\left[\mathbf{V}_1, \ldots, \mathbf{V}_N\right]$. Note that the channel dimension of each head is $C/N$. Then, we calculate the self-attention map $\mathbf{A}_j$ for each head $j$ as follows. 
\begin{equation}
\mathbf{A}_j = \mathbf{V}_j \cdot \operatorname{Softmax}(\sigma_j\mathbf{K}_j^{\top}\mathbf{Q}_j), \quad j\in\{1,2,\ldots, N\} ,
\end{equation}
where $\sigma_j$$\in$$\mathbb{R}$ is the learnable parameter to represent the variation of spectral intensity. Finally, the above output is reshaped and fused via a convolution and added with the results of position encoding $\mathcal{H}_{pos}(\cdot)$. 
\begin{equation}
    \mathbf{O}_c = \operatorname{CSA}(\mathbf{I}_c) = \left[\mathbf{A}_1, \ldots, \mathbf{A}_N\right]\mathbf{W} + \mathcal{H}_{pos}(\mathbf{V}),
\end{equation}
where $\mathbf{W}$$\in$$\mathbb{R}^{C \times C}$ is the learnable weight of the convolution. The position encoding $\mathcal{H}_{pos}(\cdot)$ is composed of several depth-wise convolutions and GELU activation. Meanwhile, inspired by the previous work \cite{zhang2022practical}, the spatial-wise transformer block (STB) aims to introduce local biases and spatial cues for the long-range transformer blocks via the sliding-window-based attention mechanism and residual blocks. Specifically, the input feature is firstly processed via an $1 \times 1$ convolution and split in the channel dimension as follows. 
\begin{equation}
\mathbf{I}_s^{(1)}, \mathbf{I}_s^{(2)} = \operatorname{Split}(\operatorname{Conv}(\mathbf{I}_s)).
\end{equation}
One part of the split feature is fed to the Swin transformer block \cite{liang2021swinir, liu2021swin} to explore the spatial correlation and non-local information. Additionally, the other part of the feature is fed to the residual block to capture local image cues. Finally, the two parts of features are fused adaptively via an $1 \times 1$ convolution as follows.
\begin{equation}
    \mathbf{O}_s^{(1)}, \mathbf{O}_s^{(2)} = \operatorname{SwinT}(\mathbf{I}_s^{(1)}), \operatorname{RB}(\mathbf{I}_s^{(2)}),
\end{equation}
\begin{equation}
    \mathbf{O}_s = \operatorname{Conv}(\left[\mathbf{O}_s^{(1)}, \mathbf{O}_s^{(2)}\right]) + \mathbf{I}_s,
\end{equation}
where $\operatorname{SwinT}(\cdot)$ and $\operatorname{RB}(\cdot)$ respectively denote the operation of sliding-window-based transformer block and residual block.

\section{Visualization Results}
\label{sec: results}
\subsection{Real Data and Simulation Results Visualization}
\begin{figure}[h]
	\centering
	\includegraphics[width=1\linewidth]{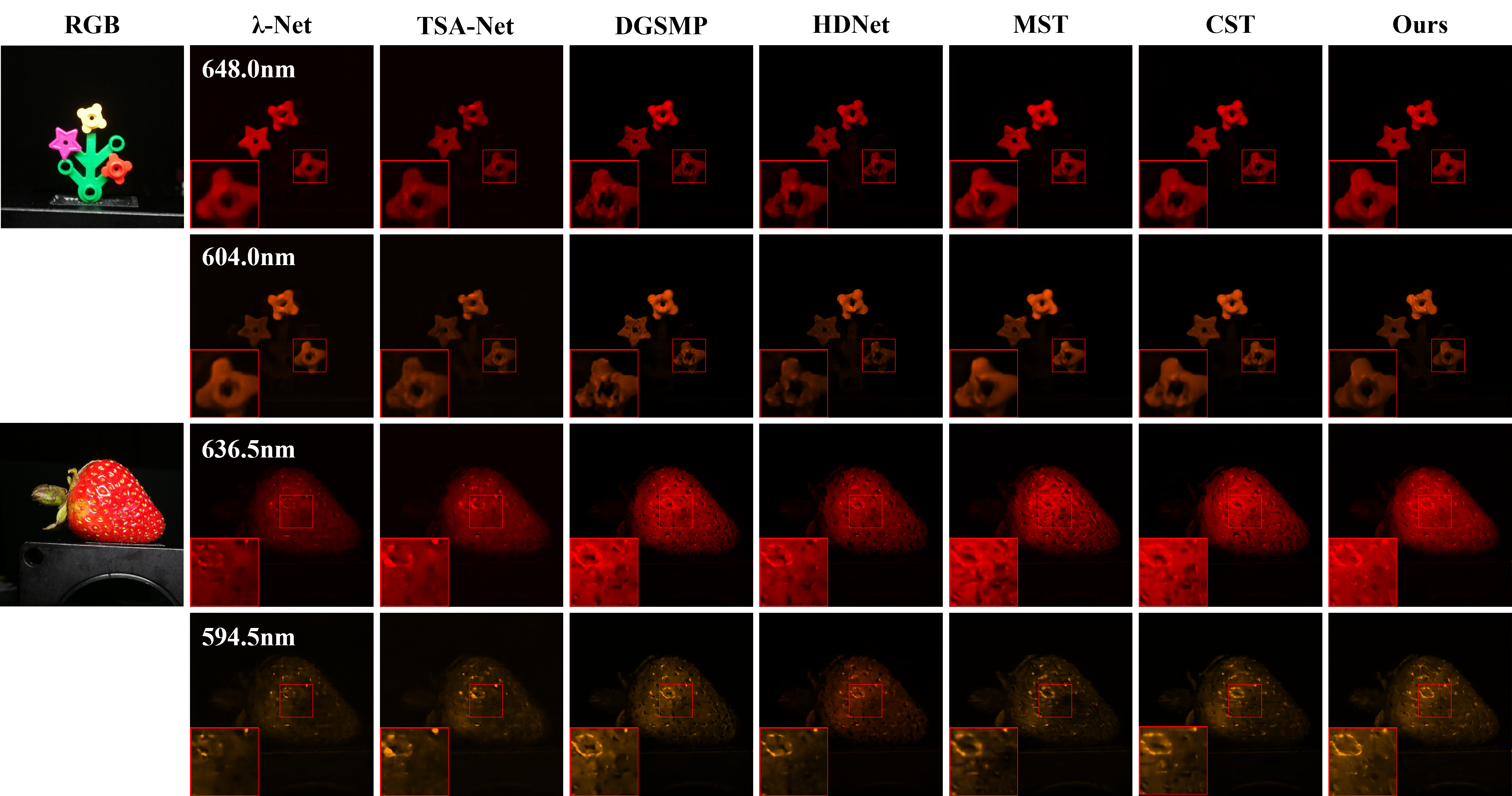}
	\caption{Real data results of the proposed RND-HRNet (Ours) and other SOTA methods on the single-shot reconstruction.}
	\label{realsup}
\end{figure}

To objectively evaluate the proposed RND-HRNet on real data, we present more visualization results with sufficient comparison methods \cite{miao2019net,meng2020end, huang2021deep, hu2022hdnet, cai2022mask, cai2022coarse} and more band numbers in 2 real measurements \cite{huang2021deep}. To simulate the real imaging situations, 11-bit shot noise is injected into the training data \cite{meng2020end}. As shown in Fig.~\ref{realsup}, our results are more perception-friendly with clearer HSI contents and fewer artifacts. For instance, our results in the cropped regions of the flowers have smoother textures and our results in the cropped regions of the strawberries have less noise, which suggests the generalization ability and robustness of our method. Meanwhile, we also illustrate the other 3 real scenes \cite{huang2021deep} in Fig.~\ref{logo} (right column). It can be clearly seen that our method can reconstruct high-quality hyperspectral images in different contents and degradations.

Furthermore, we present the two-shot simulation results of all the 28 bands \cite{huang2021deep} from 453.5nm to 648.0nm in Fig.~\ref{logo} (right column). It proves that our results have accurate spectral information and smooth spatial details in different wavelengths.
 \begin{figure}[!]
	\centering
	\includegraphics[width=1\linewidth]{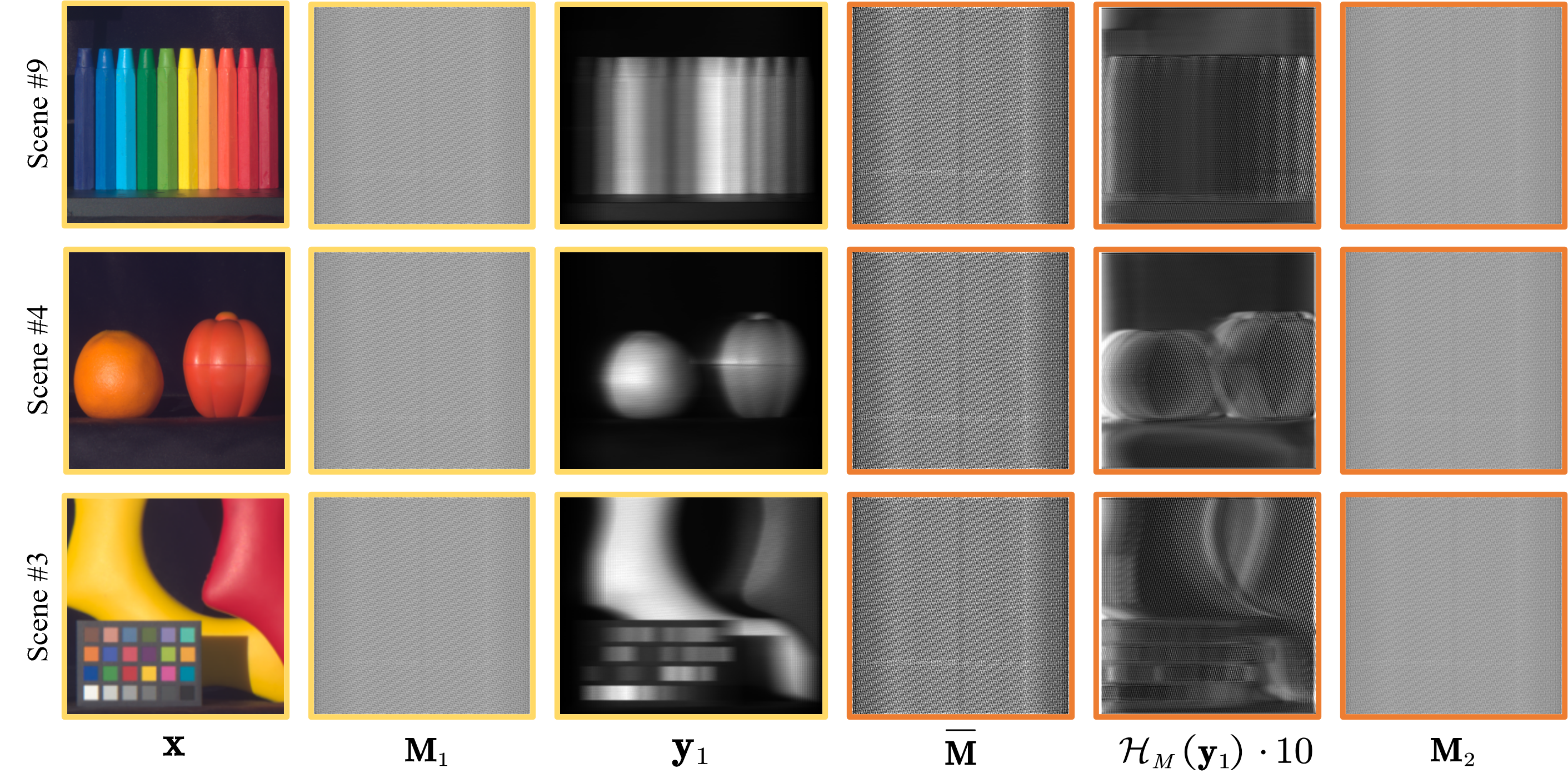}
	\vspace{-15pt}
	\caption{The visualization of the optimized content-aware masks $\mathbf{M}_1\in\mathbb{R}^{H\times W}$ and $\mathbf{M}_2\in\mathbb{R}^{H\times W}$. The content-aware mask $\mathbf{M}_2$ learns the anisotropic HSI information of different scenes adaptively with $\mathbf{M}_2=\overline{\mathbf{M}_2}+\eta^{(2)} \cdot \mathcal{H}_M^{\prime(2)}(\mathbf{y}_1)$, while the shared $\mathbf{M}_1$ and $\overline{\mathbf{M}_2}\in\mathbb{R}^{H\times W}$ learn the general characteristics of the imaging system.}
	\label{masksup}
\end{figure}
\subsection{Mask Visualization}
\label{sec: sr}
To show the structure and data distribution of optimized coded apertures in different scenarios, we visualize more content-aware masks. As shown in Fig.~\ref{masksup}, optimized masks $\mathbf{M}_1$ and $\mathbf{M}_2$ in three scenes are presented, where $\mathbf{M}_1$ is shared by all scenes and $\mathbf{M}_2$ is content-aware. Concretely, $\mathbf{M}_2$ is composed of the shared component $\overline{\mathbf{M}_2}$ and content-aware component $\mathcal{H}^{\prime(2)}_M(\mathbf{y}_1)$. We find that although $\mathbf{M}_2$ remains a similar pattern, it veils some anisotropic HSI information of different scenes, which is conducive to the multiple-shot reconstruction. 

\begin{table}[h]
	\renewcommand{\arraystretch}{1.}
	\centering
	\caption{Evaluation of the effectiveness of different components.}
	\label{tablesup}
	\resizebox{!}{1.3cm}{
		\centering
		\begin{tabular}{c|c|c|c|c|c}
			\toprule[1.5pt]
			Case Index & RNDM & CTB & STB & PSNR  & SSIM  \\ \midrule[0.75pt]
			(a)  & $\usym{2717}$   & $\usym{2713}$  & $\usym{2713}$   & 35.13 & 0.951 \\
			(b)  & $\usym{2713}$  & $\usym{2717}$    & $\usym{2713}$   & 36.15 & 0.950 \\
			(c)  & $\usym{2713}$  & $\usym{2713}$  & $\usym{2717}$     &  36.10 & 0.953 \\
			(d)  & $\usym{2713}$  & $\usym{2713}$  & $\usym{2713}$   & 36.66 & 0.957 \\ \bottomrule[1.5pt]
	\end{tabular}}
\end{table}
\section{More Ablation Studies}
\label{sec: aba}
\subsection{Ablation Study on the Proposed RND-HRNet}
To evaluate the contribution of different components in the proposed RND-HRNet, we conduct an ablation study on the single-shot reconstruction with the two-phase reconstruction network. We mainly focus on our adopted three modules, namely range-nullspace decomposition module (RNDM), channel-wise transformer block (CTB), and spatial-wise transformer block (STB). Tab.~\ref{tablesup} illustrates the PSNR (dB) and SSIM on the different settings. Removing the RNDM, we retrained a variant of the proposed model that contains only two proximal mapping modules. It can be clearly seen that the PSNR has a decline of 1.43dB, proving the effectiveness of the proposed RNDM. Meanwhile, we respectively remove the CTB and STB in the SSFM to implement two variant models, namely case (b) and case (c). Obviously, without CTB, the values of the PSNR and SSIM have dropped by 0.51dB and 0.007 respectively. Without STB, the results of PSNR and SSIM have decreased by 0.56dB and 0.004. It further proves that integrating the spectral and spatial correlation cooperatively in the SSFM is necessary and significant for reconstruction quality.

\label{sec: aba1}

\subsection{Ablation Study on the $N$-shot Reconstruction}
\label{sec: mulshot}
\begin{figure}[h]
	\centering
	\includegraphics[width=0.6\linewidth]{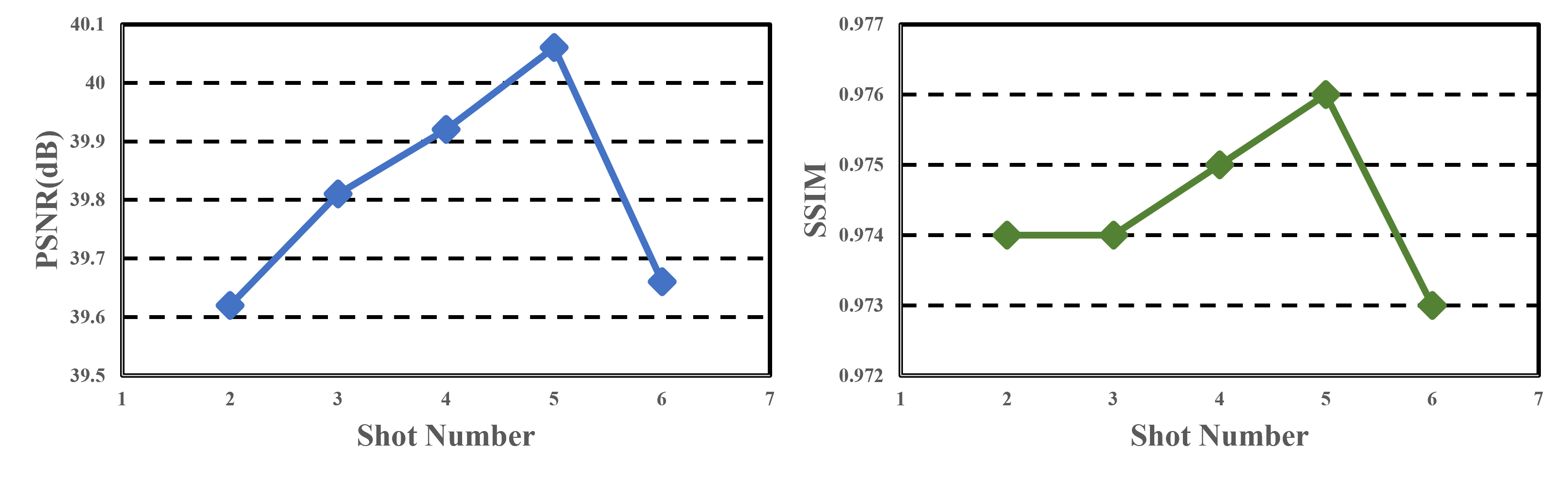}
	\caption{The values of PSNR and SSIM of the proposed PCA-CASSI on the $N$-shot reconstruction.}
	\label{abashot}
    \vspace{-10pt}
\end{figure}

\begin{figure}[h]
	\centering
	\includegraphics[width=1.\linewidth]{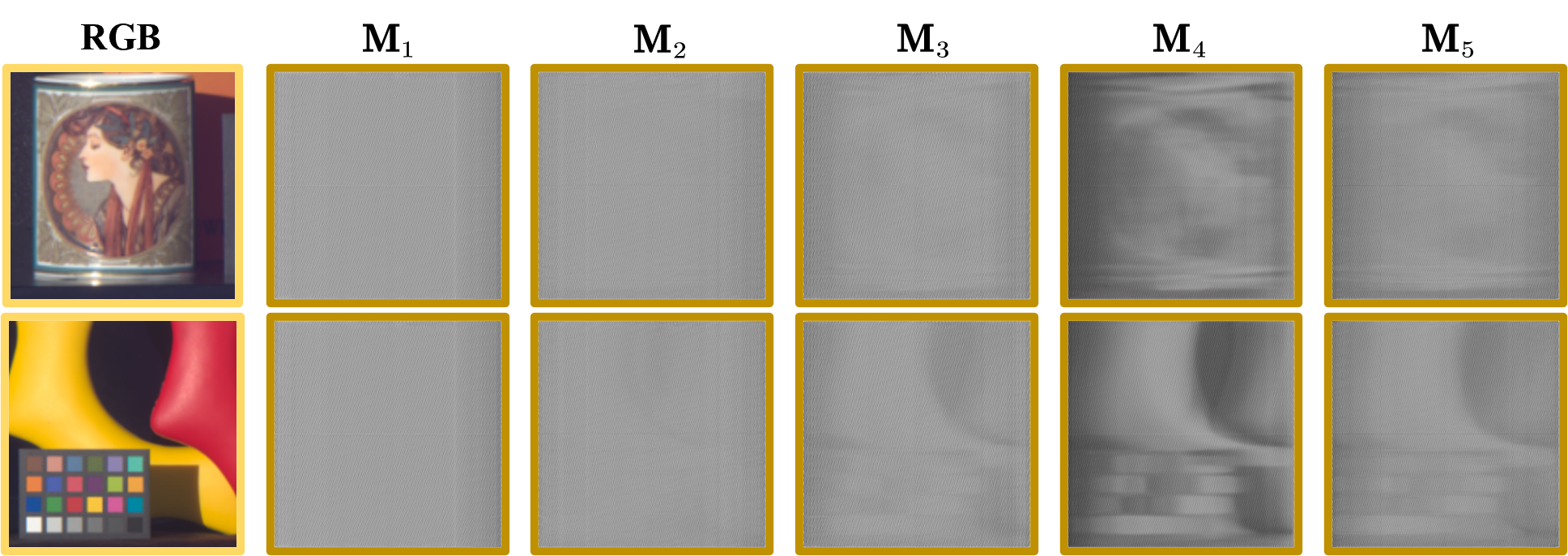}
	\caption{The visualization of five optimized content-aware masks on the $5$-shot HSI reconstruction.}
	\label{abashot1}
\end{figure}

To explore the upper bound of the proposed progressive sampling in the multiple-shot reconstruction, we retrain the proposed PCA-CASSI when the shot number $N$ is set to be 2, 3, 4, 5, and 6. The values of PSNR and SSIM are reported in Fig.~\ref{abashot}. As the shot number increases, the reconstruction performances of the proposed PCA-CASSI are improved correspondingly and reach the peak when the shot number is set to 5. However, as the shot number increases above 5, the PSNR and SSIM results decline slightly due to the increased parameters and the network overfitting. Furthermore, it can be clearly seen from Fig.~\ref{abashot1} that with the shot number increases, more HSI contents will be reflected in the coded apertures, which is conducive to the HSI reconstruction. 

\section{Limitations and Broader Impacts}
\label{sec: limit}
Limited by the quality of the hardware devices and the scarcity of real hyperspectral data, the core contribution of our paper focuses on designing a novel high-quality imaging framework and the proposed progressive sampling is mainly verified in the simulation setup, which will inspire future work in the community. However, we maintain that the proposed PCA-CASSI can be easily deployed on existing imaging systems such as \cite{wagadarikar2008single}. We will realize the proposed PCA-CASSI in hardware imaging devices and present some real data results in our future work.

The proposed PCA-CASSI contributes to the industrial application of hyperspectral compressive imaging and inspires the design of deep unfolding networks in other image inverse problems such as video snapshot compressive imaging~\cite{wt2021dense}, super-resolution~\cite{ma2021deep}, and general image restoration~\cite{mou2022deep}. Meanwhile, benefiting from its fast imaging speed and accurate HSI reconstruction, the proposed PCA-CASSI has the potential to promote closer integration of artificial intelligence in life sciences and cytology.


\end{appendices}

\end{document}